\documentclass[5p]{elsarticle}

\usepackage[utf8]{inputenc}
\usepackage[colorlinks=true,urlcolor=blue]{hyperref}
\usepackage{hyphenat}
\usepackage{amsmath, amssymb}
\usepackage{bm}
\usepackage{todonotes}
\usepackage{listings}
\usepackage{color}
\usepackage{textcomp}
\usepackage{enumerate}

\graphicspath{{figs/}} %Setting the graphicspath

\usepackage{nicefrac}
\def\onehalf{\nicefrac{1}{2}}
\makeatletter
\def\ps@pprintTitle{%
 \let\@oddhead\@empty
 \let\@evenhead\@empty
 \def\@oddfoot{}%
 \let\@evenfoot\@oddfoot}
\makeatother

\usepackage{xcolor}

\definecolor{commentsColor}{rgb}{0.497495, 0.497587, 0.497464}
\definecolor{keywordsColor}{rgb}{0.000000, 0.000000, 0.635294}
\definecolor{stringColor}{rgb}{0.558215, 0.000000, 0.135316}
\definecolor{orange}{RGB}{255,127,0}

\lstdefinestyle{mystyle}{
  basicstyle=%
    \ttfamily
    \lst@ifdisplaystyle\scriptsize\fi
}
\biboptions{sort&compress}

\makeatletter
\providecommand{\doi}[1]{%
	\begingroup
	\let\bibinfo\@secondoftwo
	\urlstyle{rm}%
	\href{http://dx.doi.org/#1}{%
		doi:\discretionary{}{}{}%
		\nolinkurl{#1}%
	}%
	\endgroup
}
\makeatother
\newcounter{bla}

\journal{Computer Physics Communications}

\begin{document}

\begin{frontmatter}

\title{A new software implementation of the Oslo method with rigorous statistical uncertainty propagation}

\author[a]{J{\o}rgen E. Midtb{\o}\corref{cor1}\fnref{first}}
\cortext[cor1]{Corresponding authors: jorgenem@gmail.com (J.E.\ Midtbø),
fabio.zeiser@fys.uio.no (F.\ Zeiser), erlenlim@fys.uio.no (E.\ Lima).}

\author[a]{Fabio Zeiser\corref{cor1}\fnref{first}}
% \cortext[cor2]{fabio.zeiser@fys.uio.no}
\fntext[sharedfirst]{Shared first authors.}

\author[a]{Erlend Lima\corref{cor1}}
% \cortext[cor3]{erlenlim@fys.uio.no}

\author[a]{Ann-Cecilie Larsen}
\author[a,b]{Gry M. Tveten}
\author[a]{Magne Guttormsen}

\author[a]{Frank Leonel Bello Garrote}
\author[a]{Anders Kvellestad}
\author[a]{Therese Renstr{\o}m}

\address[a]{Department of Physics, University of Oslo, N-0316 Oslo, Norway}
\address[b]{Expert Analytics AS, 0160 Oslo, Norway}

\begin{abstract}
The Oslo method comprises a set of analysis techniques designed to extract nuclear level density and average $\gamma$-decay strength function from a set of excitation-energy tagged $\gamma$-ray spectra. Here we present a new software implementation of the entire Oslo method, called {\ttfamily OMpy}. We provide a summary of the theoretical basis and derive the essential equations used in the Oslo method. In addition to the functionality of the original analysis code, the new implementation includes novel components such as a rigorous method to propagate uncertainties throughout all steps of the Oslo method using a Monte Carlo approach. The resulting level density and $\gamma$-ray strength function have to be normalized to auxiliary data. The normalization is performed simultaneously for both quantities, thus preserving all correlations. The software is verified by the analysis of a synthetic spectrum and compared to the results of the previous implementation, the {\ttfamily oslo\hyp{}method\hyp{}software}.

\begin{small}
\noindent
{\bf PROGRAM SUMMARY}\\
\noindent
{\em Program Title:}  \href{https://github.com/oslocyclotronlab/ompy}{{\ttfamily OMpy}} \cite{OMpyRepo2020}\\
{\em Code Ocean Capsule:}  \href{https://doi.org/10.24433/CO.6094094.v1}{{\ttfamily OMpy}} \cite{Zeiser2020b}\\
{\em Licensing provisions:} GPLv3          \\
{\em Programming language:}  Python, Cython                    \\
 {\em Reference of previous version:} \href{http://dx.doi.org/10.5281/zenodo.2318646}{{\ttfamily oslo\hyp{}method\hyp{}software}}\\
 {\em Does the new version supersede the previous version?:}  Yes   \\
 {\em Reasons for the new version:}  Facilitate modular program flow and reproducible results in a transparent and well-documented code base. Updated uncertainty quantification: formerly a stage-wise normalization without built-in uncertainty propagation. \\
 {\em Summary of revisions:} Complete reimplementation; ensemble based uncertainty quantification throughout whole method; fitting based on well-tested external libraries; corrections for the normalization procedure; auto-documentation with {\ttfamily Sphinx} \\
{\em Nature of problem:} Extraction of the nuclear level density and average $\gamma$-ray strength function from a set of excitation-energy tagged $\gamma$-ray spectra including the quantification of uncertainties and correlations of the results.    \\
  %Describe the nature of the problem here. \\
{\em Solution method:} The level density and $\gamma$-ray strength function can be obtained simultaneously using a set of analysis techniques called the Oslo method. To propagate the uncertainty from the counting statistics, we analyze an ensemble of perturbed spectra, which are created based on the experimental input. One obtains a set of level densities and $\gamma$-ray strength functions for each realization from a fit process. The fitting metric ($\chi^2$) is degenerate, but the degeneracy is removed by a \emph{simultaneous} normalization of the level density and $\gamma$-ray strength function to external data, such that all correlations are preserved.\\
  %Describe the method solution here.
%{\em Additional comments including Restrictions and Unusual features (approx. 50-250 words):}\\
  %Provide any additional comments here.
\end{small}

\end{abstract}

\begin{keyword}
Oslo method; Nuclear level density; $\gamma$-ray strength function; Degenerate inverse problem
\end{keyword}

\end{frontmatter}
\section{Introduction}
One long-standing challenge in nuclear physics is to precisely determine nuclear properties at excitation energies above the discrete region
 % ($\simeq 2$ MeV)
and up to the particle threshold(s).
This region is often referred to as the \emph{quasicontinuum} and represents an excitation-energy range where the quantum levels are very closely spaced. That leads to a significant degree of mixing (complexity) of their wave functions, but they are still not fully overlapping as in the continuum region.
For the quasicontinuum, it is fruitful to introduce average quantities to describe the excited nucleus: instead of specific levels, the level density $\rho(E_x)$ as a function of excitation energy $E_x$ is used, and instead of specific reduced transitions strengths $B(XL)$ between a given initial and final state, the average decay strength represented by the $\gamma$-ray strength function ($\gamma$SF) is applied.

In addition to their key role in describing fundamental nuclear properties, both the level density and the $ \gamma$-ray strength function are vital components for calculating cross sections and reaction rates for explaining the nucleosynthesis of heavy elements in astrophysics \cite{Arnould2007, Larsen2019}. The ability to calculate cross sections may also help in the design of next generation nuclear reactors, where direct cross section measurements are  missing or have too high uncertainties for the given application~\cite{NEAHPPuNG32, NEAHPPu33, Salvatores2008, Harada2014}.

The Oslo method \cite{Rekstad1983, Henden1995, Tveter1996, Schiller2000} allows for extracting the level density $\rho$ and the $\gamma$SF simultaneously from a data set of particle $\gamma$-ray coincidences. It has been implemented in a collection of programs known as the {\ttfamily oslo\hyp{}method\hyp{}software}~\cite{Oslo-v1.1.2}, and the analysis has been successfully applied to a range of nuclei in widely different mass regions \cite{Larsen2018,Renstrom2018,Renstroem2016,Laplace2015}.
However, the Oslo method consists of several non-linear steps.
This makes an analytical propagation of statistical and systematic uncertainties very difficult, and thus hampers a reliable uncertainty quantification for the final results. The statistical uncertainty propagation from unfolding the $\gamma$ spectra and the determination of the primary $\gamma$-ray distribution has so far not been addressed in a fully rigorous way. In lieu of this, an approximate uncertainty estimation has been used, which is described in Ref.\ \cite{Schiller2000}. Moreover, uncertainties related to the absolute normalization of the level density $\rho$ and $\gamma$SF have been discussed in Ref.\ \cite{Larsen2011}, but there was no automatized way to include this in the final results. Approximate ways to include normalization uncertainties have been suggested and used, see e.g.\ Ref.\ \cite{Kullmann2019}, but they do not account for correlations between parameters as they were not available within the {\ttfamily oslo\hyp{}method\hyp{}software}.

In this work, we approach the problem of uncertainty propagation using a Monte Carlo technique. By generating an ensemble of perturbed input spectra, distributed according to the experimental uncertainties, and propagating each ensemble member through the Oslo method, we can gauge the impact of the count statistics on the final results. The resulting level density and $\gamma$-ray strength function have to be normalized to external data. We have implemented a new simultaneous normalization for both quantities, thus preserving all correlations between them.

In the following, we present {\ttfamily OMpy}, the new implementation of the Oslo method. We discuss the various steps of the method and present our new uncertainty propagation and normalization routine. The code is tested by the analysis of a synthetic spectrum. The capability of the new method is illustrated by applying it to experimental data and a comparison to the previous implementation.

\section{Usage of {\ttfamily OMpy}}

{\ttfamily OMpy} is designed to facilitate a more complete uncertainty propagation through the whole Oslo method and at the same time simplify the user interface and enhance the reproducibility of the analysis. This section will focus on the latter two goals. \href{https://github.com/oslocyclotronlab/ompy}{{\ttfamily OMpy}}, as well as all the  \href{https://doi.org/10.24433/CO.6094094.v1}{Jupyter notebooks and datasets} used to create the figures in this article, is available online \cite{OMpyRepo2020, Zeiser2020b}.

The documentation of the interface of {\ttfamily OMpy} is created automatically from its source code with the \href{https://sphinx-automodapi.readthedocs.io}{\ttfamily sphinx\hyp{}auto\-mod\-api} package and is available from \href{https://ompy.readthedocs.io/}{ompy.readthedocs.io}. A detailed example of the usage is provided in the {\ttfamily getting\-started} Jupyter notebook. Taking a step beyond just listing the version number, we have set up a {\ttfamily docker} container to ensure also the reproducibility of the software environment in which the user runs the analysis. The notebooks can be run interactively online without installation through Binder \cite{Jupyter2018}, although a limit on the available computation power can lead to considerable computation times for cases with large ensembles. For larger calculations we recommend the Code Ocean [DOI follows with publication] capsule.

As most operations within the Oslo method will require working on binned quantities like $\gamma$-ray spectra or level densities, or a collection of spectra into matrices, we have created the {\ttfamily Vector} and {\ttfamily Matrix} classes. These store the count data as one- or two-dimensional {\ttfamily NumPy} \cite{Oliphant2006,Walt2011} arrays, together with an array giving the energy calibration for each axis. The classes also contain a collection of convenience functions e.g.\ for saving and loading files to disk, rebinning and plotting.

\section{The Oslo method}
\label{sec:oslo-method}

\begin{figure*}[bt]
\begin{center}
\includegraphics[clip,width=2.\columnwidth]{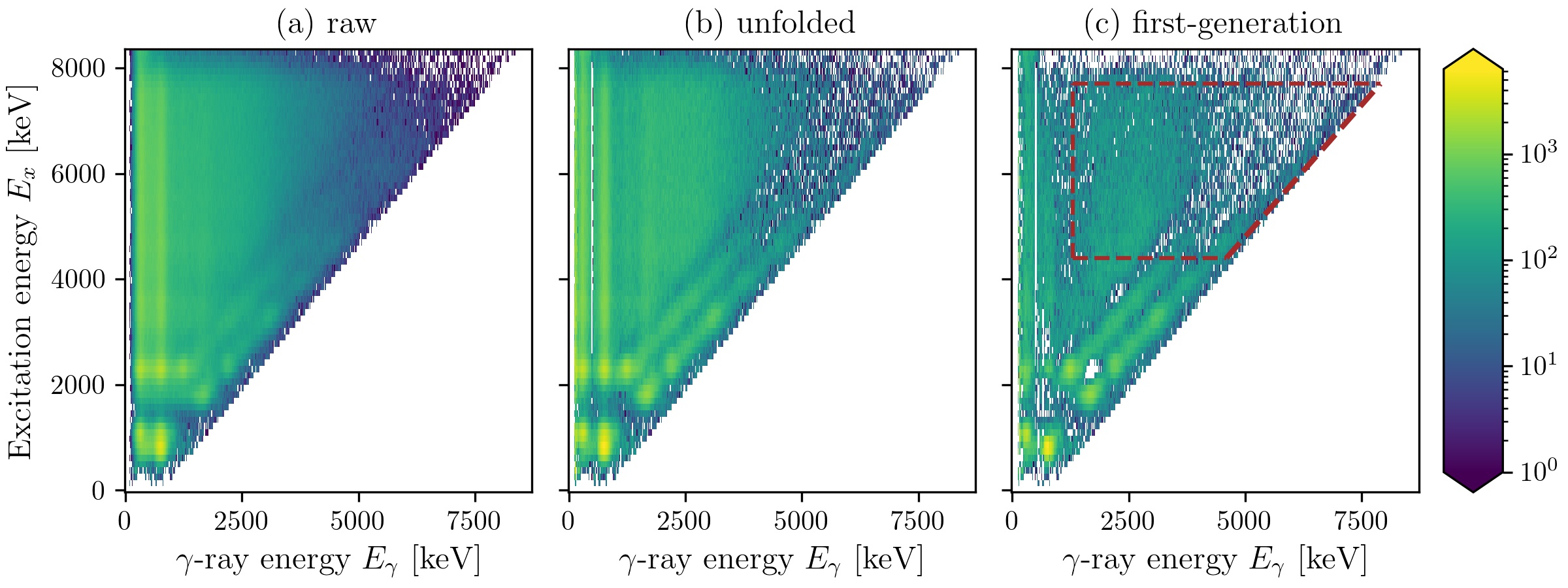}
\includegraphics[clip,width=2.\columnwidth]{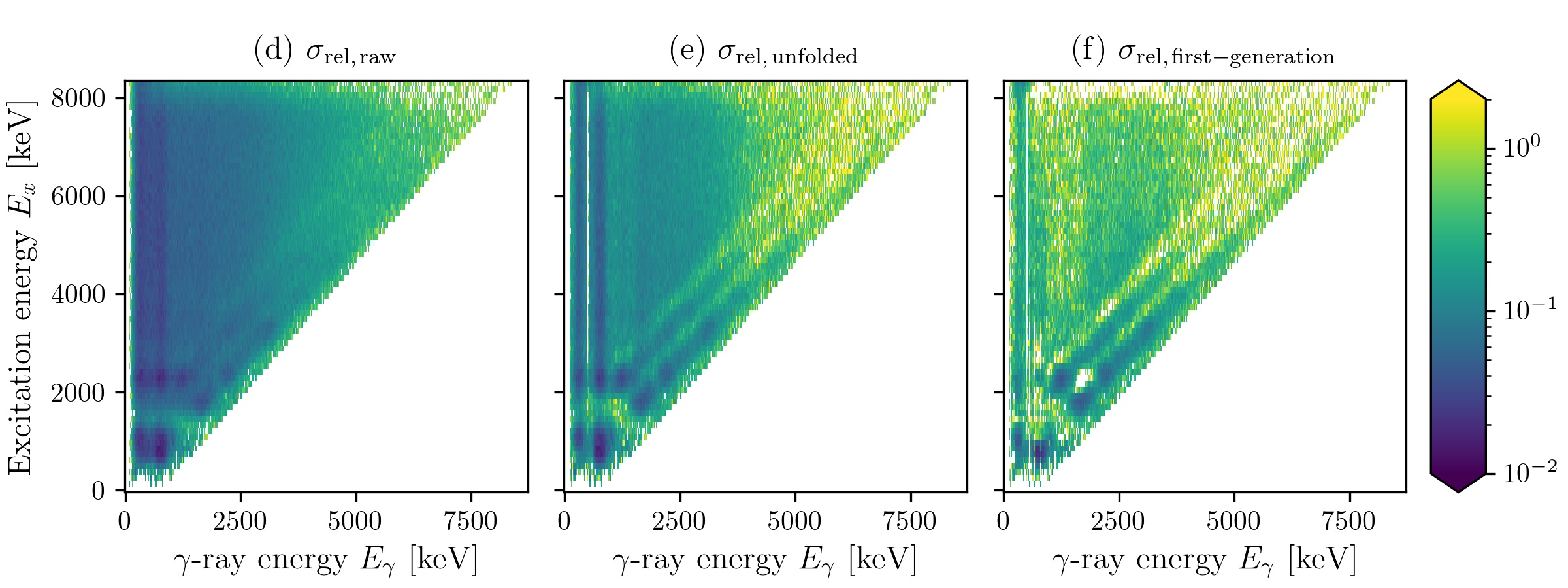}
%\vskip 2cm
\caption{Raw (a), unfolded (b) and first-generation (c) matrices for the $^{164}$Dy dataset \cite{Nyhus2010,Renstrom2018}, as well as the respective relative standard deviation matrices (d--f) obtained with the ensemble propagation technique for $N_\mathrm{ens}=50$ realizations. The dashed lines in panel (c) highlight the fit limits for $\rho$ and $f$.}
\label{fig:Dy164_matrices}
\end{center}
\end{figure*}

The starting point for the Oslo method is an $E_x$--$E_\gamma$ \emph{coincidence matrix}, i.e.\ a set of $\gamma$-ray spectra each stemming from an identified initial excitation energy $E_x$.
A standard way to construct this input matrix is from coincidence measurements of $\gamma$ rays and charged ejectiles following inelastic scattering or transfer reactions with light ions. An array of $\gamma$-ray detectors measures the energy $E_\gamma$ of the emitted $\gamma$s, while a particle telescope determines the excitation energy $E_x$ using the energy deposited from the outgoing charged particles. (For a detailed description, see e.g.\ Ref.\ \cite{Larsen2011} and references therein.) Recently, other methods have been developed that obtain the coincidence matrix from $\gamma$ rays in $\beta$-decay \cite{Spyrou2014} or in an inverse kinematics setup \cite{Ingeberg2020}.
An example of a coincidence matrix for $^{164}$Dy from a standard Oslo method experiment \cite{Nyhus2010,Renstrom2018} at the Oslo Cyclotron Laboratory is shown in panel (a) of Fig.\ \ref{fig:Dy164_matrices}.
%---------------------------------------------------%

%%%%%%%%%%%%
% Unfolding
%%%%%%%%%%%%
The first step of the Oslo method is to \emph{unfold}, i.e., deconvolute the $\gamma$-ray spectra for each excitation energy to compensate for the detector response (Compton scattering, pair production, etc.).
This is implemented in the {\ttfamily Unfolder} class using an iterative unfolding method described in Ref.\ \cite{Guttormsen1996}. We reiterate the main points of the procedure in \ref{app:unfolding}. There we also describe a routine to select the best iteration, which has already been implemented in the {\ttfamily oslo\hyp{}method\hyp{}software} but not yet published elsewhere. The unfolded $^{164}$Dy spectrum is shown in Fig.\ \ref{fig:Dy164_matrices} (b).

%%%%%%%%%%%%
% First-generation
%%%%%%%%%%%%
The second step is the determination of the \textit{first\hyp{}generation}, or \emph{primary}, $\gamma$-ray spectrum for each excitation energy.
In the {\ttfamily FirstGeneration} class an iterative procedure is applied as described in Ref.\ \cite{Guttormsen1987}. We recapitulate the procedure in \ref{app:firstgen}, including a small addition to minimize fluctuations in higher order iterations. The resulting first-generation $\gamma$-ray matrix is shown in panel (c) of Fig.\ \ref{fig:Dy164_matrices}. The main assumption of the first-generation method is that the $\gamma$-ray spectra following the populations of levels within the excitation energy bin $E_x$ by the inelastic or transfer reaction are the same as when this excitation energy bin is populated via $\gamma$-ray decay from higher lying levels \cite{Larsen2011}. Although this is plausible at first sight, it requires that the spin-parity distribution of the populated levels is approximately independent of the excitation energy \cite{Larsen2011, Zeiser2018a}.

%%%%%%%%%%%%
% Decomposition
%%%%%%%%%%%%
The next step of the Oslo method consists of fitting the first-generation spectra by a product of two one\hyp{}dimensional functions, namely the \emph{nuclear level density} $\rho(E_x)$ and the \emph{$\gamma$-ray transmission coefficient} $\mathcal{T}(E_\gamma)$.
The method relies on the relation
 \begin{align}
  P(E_x, E_\gamma) \propto \rho(E_f = E_x - E_\gamma) \mathcal{T}(E_\gamma),\label{eq:Oslo_method_eq}
\end{align}
where we look at deexcitations from an initial excitation energy bin $E_x$ to the final bin $E_f$, such that the $\gamma$-ray energy is $E_\gamma=E_x - E_f$. Here, $P(E_x, E_\gamma)$ is a matrix of first-generation spectra $FG(E_\gamma)_{E_x}$ normalized to unity for each $E_x$ bin.\footnote{Note that we follow the standard notation for the Oslo method, where $P(E_x, E_\gamma)$ is the conditional probability $p(E_\gamma | E_x )$ for the first $\gamma$-ray transition with energy $E_\gamma$ to come from an initial excitation energy $E_x$.}
Furthermore, if we assume that the $\gamma$ decay at high $E_x$ is dominated by dipole radiation, the transmission coefficient is related to the dipole \emph{$\gamma$-ray strength function} $f(E_\gamma)$ (or $\gamma$SF) by the relation
\begin{align}
    \mathcal{T}(E_\gamma) = 2\pi E_\gamma^3 f(E_\gamma).\label{eq:gammaSF}
\end{align}
A derivation of Eq.\ \eqref{eq:Oslo_method_eq} is shown in \ref{app:oslomethodeq}, where the main assumptions underpinning this decomposition are:
\begin{itemize}
    \item The compound nucleus picture: We assume that the $\gamma$ decay from an excited nuclear state is independent of how the excited state was formed. This goes back to Bohr's theory for compound nuclei \cite{Bohr1936} and is supported by many experiments \cite{Wallner1995,Voinov2007,Voinov2009,Ramirez2015,Sherr1961,Zhuravlev2011,Renstrom2018,Guttormsen2005,Larsen2017}.
    \item Dominance of dipole radiation: It is assumed that the decay is dominated by dipole radiation. This is strongly supported by data and theory, see e.g.\ Refs.\ \cite{Bassauer2016, Kinsey1954, Kopecky1981, Kopecky1987, Chrien1984, Winter1989, Vennink1980, Jones2018, Rusev2013, Larsen2013, Larsen2017,Simon2016}.
    \item The generalized Brink-Axel hypothesis: The $\gamma$-ray strength function $f(E_\gamma)$ is independent of the initial and final states, i.e., it is the same for excitations and decays between any initial and final states that are separated by the energy $E_\gamma$ \cite{Brink1955,Axel1962,Guttormsen2016,Campo2018,Becvar1992,Krticka2004,Kopecky1987,Johnson2015}.
    \item The population cross-section: To obtain the ``summed'' level density $\rho(E_f)=\sum_J \rho(E_f, J)$ in Eq. \eqref{eq:Oslo_method_eq} the population cross-section has to be approximately proportional to the intrinsic spin distribution $g(E_x, J)$ (and we assume parity equilibration, i.e.\ $\rho(E_x, J, \pi=+) \approx \rho(E_x, J, \pi=-)$). When the employed reaction is very spin selective, like in $\beta$-decay, a weighted sum of the level densities is obtained instead, see Eq.\ \eqref{eq:oslo eq weighted}.
\end{itemize}

Finally, we have to normalize the nuclear level density $\rho(E_x)$ and $\gamma$-ray strength function $f(E_\gamma)$ (or equivalently, the $\gamma$-ray transmission coefficient $\mathcal{T}(E_\gamma)$). It was shown by~\citet{Schiller2000} that the solution to Eq.\ \eqref{eq:Oslo_method_eq} is invariant under a Lie group $G$ of transformations by three continuous real valued parameters $A, B$ and $\alpha$:
\begin{align}
    \rho(E_f=E_x - E_\gamma), f(E_\gamma) \stackrel{G}{\to} Ae^{\alpha E_f}\rho(E_f), B e^{\alpha E_\gamma} f(E_\gamma).
    \label{eq:G}
\end{align}
However, we stress that the degrees of freedom are limited to those given by $G$ --- i.e., all shape features of $\rho$ and $f$ beyond the exponential prefactor given by $G$ are uniquely determined by the fit. It is important to note that the $\alpha$ parameter, which influences the slope (in a log plot) of the functions, is common to $\rho$ and $f$. Hence, their normalizations are coupled together in the Oslo method. To obtain the physical solution, the level density and $\gamma$SF need to be normalized to auxiliary data. Typically, one uses $s$-wave resonance spacings, $D_0$, from neutron capture experiments as well as discrete levels to fix the level density normalization, and augment this by average total radiative width data, $\langle \Gamma_\gamma \rangle$, to normalize the $\gamma$SF. This will be discussed in more detail in the next sections.

\section{Uncertainty propagation by ensemble}
We use an approach based on the Monte Carlo (MC) technique to estimate the statistical uncertainties in the Oslo method by creating an ensemble of randomly perturbed copies of the data set under study. To illustrate the method, we have chosen an experimental data set for $^{164}$Dy. The data was obtained in Ref.\ \cite{Nyhus2010} and we will compare our results to the reanalysis published by \citet{Renstrom2018}.

The random variables are the experimental number of counts in each  bin $i$ in the raw $E_x$--$E_\gamma$ coincidence matrix $R$. We assume that they are independent and follow a Poisson distribution with parameter $\lambda_i$. The Poisson distribution $\mathcal{P}_\lambda$ is given as
\begin{align}
    \mathcal{P}_\lambda = p(k|\lambda) = \frac{\lambda^k e^{-\lambda}}{k!}
\end{align}
We take the number of counts $k_i$ in bin $i$ of $R$ as an estimate for the Poisson parameter $\lambda_i$. Note that it is an unbiased estimator for $\lambda_i$, since the expectation value $\langle k \rangle = \lambda$. To generate a member matrix $R_l$ of the MC ensemble, the counts in each bin $i$ are replaced by a random draw from the distribution $\mathcal{P}_{k_i}$. By this procedure, we obtain $N_{\mathrm{ens}}$ matrices representing different realizations of the experiment. Defining $\vec r_i = (r_i^{(1)}, r_i^{(2)}, \dots, r_i^{(N_\mathrm{ens})})^T$ as the vector of all $N_\mathrm{ens}$ realizations $m$ of bin $i$ we can calculate the sample mean $\langle \vec r_i \rangle$,
\begin{align}
    \langle \vec r_i \rangle = \frac{1}{N_\mathrm{ens}} \sum_{l=1}^{N_\mathrm{ens}} r_i^{(m)},
\end{align}
and standard deviation $\sigma_i$,
\begin{align}
    \sigma_i &= \sqrt{\frac{1}{N_\mathrm{ens}} \sum_{l=1}^{N_\mathrm{ens}} \left( r_i^{(m)} - \langle \vec r_i \rangle \right)^2}.
\end{align}
Of course, in the case of the raw matrix $R$, the standard deviation is trivial because it is given by the Poisson distribution ($\sigma = \sqrt{\lambda}$). But the technique also allows us to estimate the standard deviation at later stages in the Oslo method --- after unfolding, after the first-generation method and even after fitting the level density and $\gamma$-ray transmission coefficient.
In Fig.\ \ref{fig:Dy164_matrices} we show the relative standard deviations $\sigma_{\mathrm{rel}, i} = \sigma_i/r_i$ in the raw (d), unfolded (e) and first-generation (f) matrices of the $^{164}$Dy dataset based on $N_\mathrm{ens}=50$ ensemble members.

It should be noted that we usually analyze spectra with a (possibly time-dependent) background. In this case we measure two raw spectra, the total and the background spectra, which independently follow a Poisson distribution. In {\ttfamily OMpy}, both spectra can be read in and perturbed independently. The background subtracted spectra $R'$ are then generated for each realization. When the number of background counts are large enough, this may lead to bins in $R'$ with negative number of counts. With the current default, these are removed before further processing at the cost of a potential bias towards higher level densities $\rho$ and strength functions $f$ (see the discussion in Sec.\ \ref{sec:systematic_uncertainties}). The generation of the all ensembles matrices (including application of the unfolding and first-generation method) is handled by the {\ttfamily Ensemble} class.

\section{Decomposition into level density and $\gamma$-ray \\ strength function}
\label{sec:decomposition}
With the first-generation matrices and their standard deviation at hand, we may proceed with the fitting of $\rho$ and $\mathcal{T}$ for each ensemble member. First, we select a suitable bin size $\Delta E$, typically 100--300 keV depending on the detector resolution, and rebin the first-generation matrix along both the $E_x$ and $E_\gamma$ axes to this bin size. The matrix of experimental decay probabilities $P_\mathrm{exp}(E_x, E_\gamma)$ is obtained by normalizing the spectrum in each $E_x$ bin to unity. For the fit of $\rho$ and $\mathcal{T}$, we take the function value in each bin as a free parameter. For a given pair of trial functions $(\rho, \mathcal{T})$, the corresponding matrix $P_\mathrm{fit}(E_x, E_\gamma)$ is constructed by
\begin{align}
    \label{eq:Pfit_N_rho_T}
    P_\mathrm{fit}(E_x, E_\gamma) = C_{E_x} \rho(E_x-E_\gamma) \mathcal{T}(E_\gamma),
\end{align}
where $C_{E_x}$ is a normalization coefficient so that \linebreak[4] $\sum_{E_\gamma}P_\mathrm{fit}(E_x, E_\gamma) = 1 \,\,\forall E_x$. We fit $P_\mathrm{fit}$ by a $\chi^2$ minimization approach, minimizing the weighted sum-of-squared errors
\begin{align}
    \label{eq:chi2-pfit}
    \chi^2 = \sum_{E_x, E_\gamma} \left( \frac{P_\mathrm{exp}(E_x, E_\gamma) - P_\mathrm{fit}(E_x, E_\gamma) }{\sigma_{P_\mathrm{exp}}(E_x, E_\gamma)}\right)^2.
\end{align}
It is important to use a weighted sum rather than simply a sum of the residuals, to suppress the influence of bins with large uncertainties. This in turn makes uncertainty estimation important. As already mentioned, a shortcoming of the original Oslo method implementation~\cite{Schiller2000} in the {\ttfamily oslo\hyp{}method\hyp{}software} has been the estimation of the uncertainty $\sigma_{P_\mathrm{exp}}(E_x, E_\gamma)$ in the denominator of the $\chi^2$ fit. Due to the lack of a complete statistical uncertainty propagation, one has had to resort to an approximate uncertainty estimation. It was based on a Monte Carlo scheme similar to the present work, but where only the first-generation spectrum is perturbed as if each entry was direct count data. This is discussed in detail in Ref.\ \cite{Schiller2000}. In {\ttfamily OMpy}, we have access to a proper uncertainty matrix $\sigma_{P_\mathrm{exp}}$. We checked that most bins of the first-generation matrices approximately follow a normal distribution. However, they are distributed with a larger and varying standard deviation as compared to what one would have received if the first generation entries were count data directly and followed the expectation value $\langle k \rangle = \lambda$. The approximate adherence to the normal distribution justifies the use of a $\chi^2$ minimization as a likelihood maximization.

The $\chi^2$ minimization is carried out by numerical minimization in the {\ttfamily Extractor} class. This is a different implementation than in the {\ttfamily oslo\hyp{}method\hyp{}software}, where the minimum is found by iteratively solving a set of equations to obtain a solution satisfying $\partial \chi^2/\partial \rho = 0$, $\partial \chi^2/\partial \mathcal{T} = 0$ for each bin of $\rho$ and $\mathcal{T}$ \cite{Schiller2000}. After testing several off-the-shelf minimizers, we have found that the modified Powell's method in the {\ttfamily SciPy} package works well \cite{Jones2001,Powell1964}.

\begin{figure}[bt]
\begin{center}
\includegraphics[clip,width=1.\columnwidth]{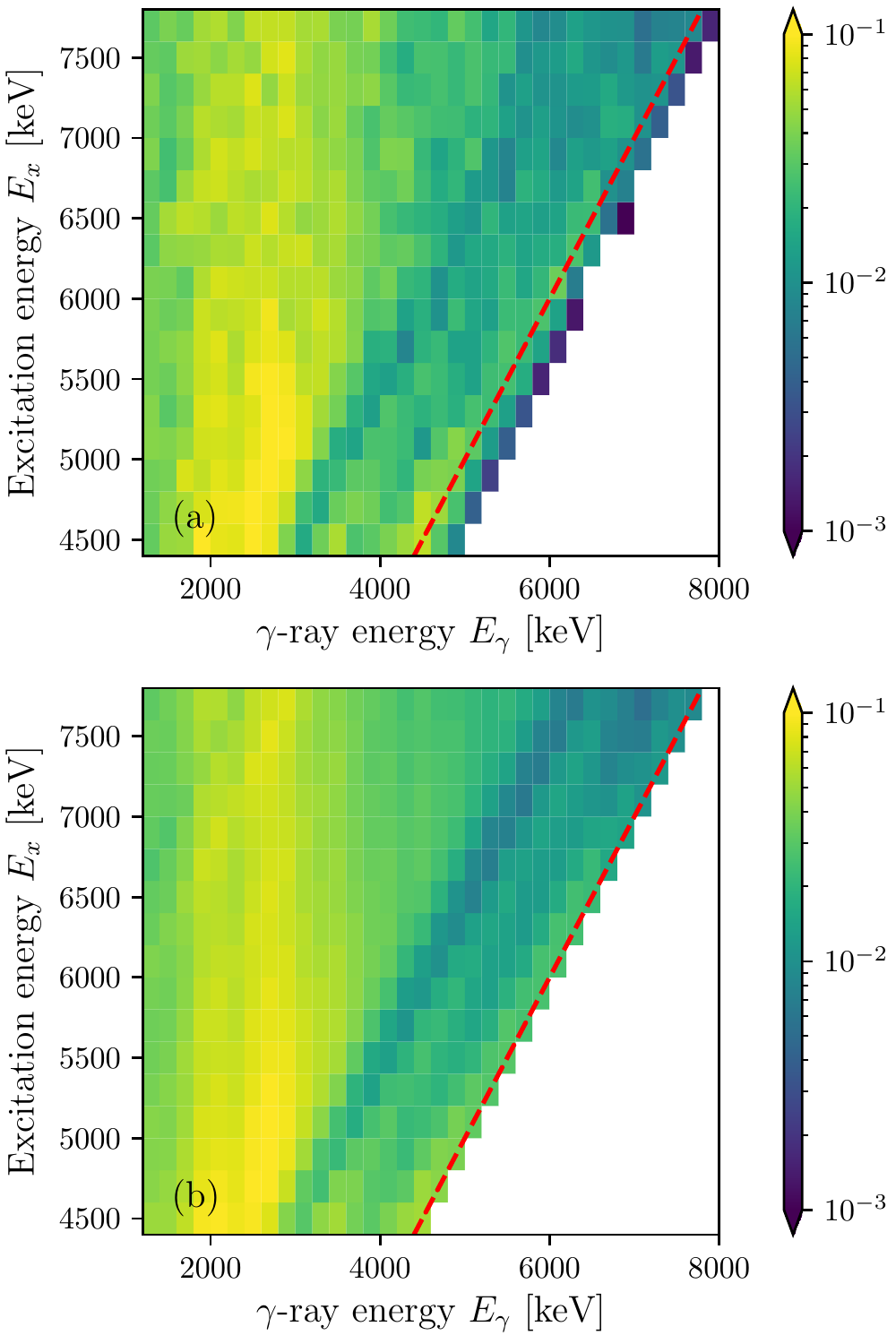}
%\vskip 2cm
\caption{One realization of the normalized first-generation matrix $P(E_x, E_\gamma)$ (a) compared to its fit (b) by the product of the level density $\rho(E_x)$ and $\gamma$SF $f(E_\gamma)$ (see Fig.\ \ref{fig:Dy164_fit_raw}). The dashed line indicates the maximum $\gamma$-ray energy $E_\gamma = E_x$. Counts to the right of this diagonal are due to the detector resolution or noise only and have been excluded from the fit. Note that panel (a) is similar to Fig.\ \ref{fig:Dy164_matrices}c, but rebinned to 200 keV and for one realization instead of the mean of the ensemble.}
\label{fig:Dy164_fg_versus_product}
\end{center}
\end{figure}

The normalized first-generation matrix $P(E_x, E_\gamma)$ is compared to the fitted matrix in Fig.\ \ref{fig:Dy164_fg_versus_product} for one ensemble member of the $^{164}$Dy dataset. In Fig.\ \ref{fig:Dy164_fit_raw} we show the corresponding level density $\rho$ and  $\gamma$-ray strength function $f$, where the latter is obtained from $\mathcal{T}$ using Eq.\ \eqref{eq:gammaSF}.

\subsection{Testing the sensitivity on the initial values}
\label{sec:decomposition_initial_guess}
The presently used minimization routine requires an initial guess for the trial functions $(\rho, \mathcal{T})$. In principle, the choice of the trial functions may have a significant effect on the results if the minimizer is prone to get caught in local minima. Ref.\ \cite{Schiller2000} proposes to set the initial $\rho$ to a constant, arbitrarily chosen as 1, for all excitation energies $E_x$, and $\mathcal{T}(E_\gamma)$ as the projection of $P_\mathrm{exp}(E_x, E_\gamma)$ on the $E_\gamma$-axis. In {\ttfamily OMpy}, we have implemented several choices to test the stability of the solutions towards other initial guesses. For the cases shown here, no significant impact on the final results was observed.

Our default choice for the initial guess on $\rho$ has been motivated by a long lasting discussion on whether the level density $\rho$ follows the back-shifted Fermi gas (BSFG) equation \cite{Bethe1936, Gilbert1965} or a constant temperature (CT) model \cite{Ericson1959, Gilbert1965} below the neutron separation energy $S_n$ , see Ref.\ \cite{RIPL3, Guttormsen2013} and references therein. In recent years, the results of the Oslo method have strongly suggested a close-to CT-behavior. As this is equivalent to the initial guess for $\rho$ in  Ref.\ \cite{Schiller2000} (after a transformation $G$ (see Eq.\ \eqref{eq:G}), we set the default initial guess to a BSFG-like solution. More specifically, we draw from a uniform distribution centered around a BSFG-like initial guess for each ensemble member. Fig.\ \ref{fig:Dy164_fit_raw} demonstrates that the resulting fit is still best described by the CT-model, but it is now very unlikely that this is due to a failure of the minimizer. The initial guess for $\mathcal{T}$ is still chosen as in Ref.\ \cite{Schiller2000}, but with the same randomization with a uniform distribution as for the level density $\rho$.

Besides the two alternatives named above, we have also tested a rather exotic initial guess for $\rho$ given by a quadratic function with a negative coefficient for the second-degree. This contrasts any expectation that the level density $\rho$ increases as a function of the excitation energy $E_x$. However, even for this choice, the solution is stable.

\subsection{Uncertainty estimation}
\label{sec:decomposition_uncertainty}
Given the degeneracy of the $\chi^2$, it is nontrivial to estimate the uncertainty in the solutions of $\rho$ and $\mathcal{T}$. The {\ttfamily oslo\hyp{}method\hyp{}software} implements the approach of Ref.\ \cite{Schiller2000}, where the uncertainty is estimated from the standard deviation of the solutions ($\rho^{(m)}$,~$\mathcal{T}^{(m)}$) for each realization of the first-generation matrix $P_\mathrm{exp}^{(m)}$. This may lead to erroneously high uncertainties, as any transformation $G$ gives an equally good solution to a given $P_\mathrm{exp}^{(m)}$. To illustrate this, we could imagine that the $P_\mathrm{exp}^{(m)}$'s only differ by a very small noise term. Due to the noise term the solutions will not be identical. In addition, we recapitulate that any given $P_\mathrm{exp}^{(m)}$ can be equally well fit by $\rho$ as with any allowed transformation $G$ such as 10$\rho$ or 100$\rho$. Instead of a negligible standard deviation one receives a standard deviation solely based on the degeneracy of the solutions. So far, we have observed that the minimization results are more stable as one might expect for the scenario outlined here (provided that the initial guess is not randomized). This is probably due to the way the minimizer explores the parameter space. Nevertheless, for the standard usage of {\ttfamily OMpy} we provide the functionality to estimate the uncertainty of solutions $\rho$ and $\mathcal{T}$ after normalization of each ensemble of ($\rho^{(m)}$,~$\mathcal{T}^{(m)}$). This will be explained in more detail in the next section.

\subsection{Fit range}
\label{sec:decomposition_fit_range}
Using the Oslo method we have to restrict the fit range for the $P(E_x, E_\gamma)$ matrix. The area below $E_{x}^\mathrm{min}$ exhibits discrete transitions, thus does not adhere to the statistical nature of the $\gamma$-ray strength function $f$ and is therefore excluded. To remain selective on the $\gamma$-decay channel, we can only use $P(E_x, E_\gamma)$ up to a maximum excitation energy $E_{x}^\mathrm{max}$ around the neutron separation energy $S_n$. Finally, we also constrain the minimum $\gamma$-ray energy to $E_{\gamma}^ \mathrm{min}$, which usually attributed to a  deficiency of the unfolding or first-generation method for low $\gamma$-ray energies. The limits are highlighted in Fig.\ \ref{fig:Dy164_matrices} (c) and only this valid region of $P(E_x, E_\gamma)$ is sent to the minimizer.

\begin{figure}[bt]
\begin{center}
\includegraphics[clip,width=1.\columnwidth]{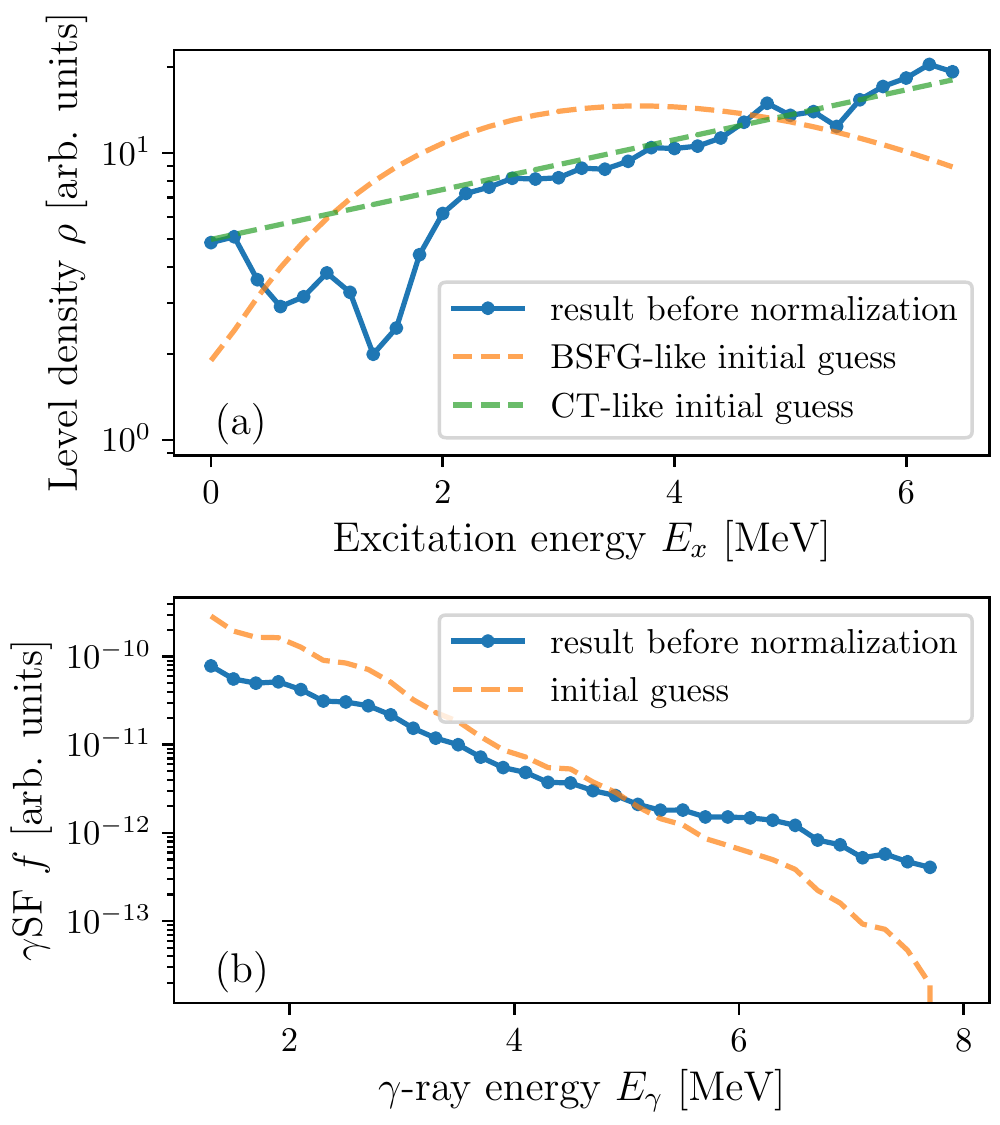}
%\vskip 2cm
\caption{Level density $\rho(E_x)$ and $\gamma$SF $f(E_\gamma) = \mathcal{T}(E_\gamma)/(2\pi E_\gamma^3)$ from the fit in Fig.\ \ref{fig:Dy164_fg_versus_product}. No transformation has been applied to the fit. Even though the initial guess for $\rho$ was chosen from a BSFG-like function, the results are better described by the CT model above $\sim 2$~MeV. The initial guess is shown before the randomization with a uniform distribution.}
\label{fig:Dy164_fit_raw}
\end{center}
\end{figure}

\section{Normalization of $\rho$ and $\gamma$SF}
At first glance, the results bear little resemblance to a level density or $\gamma$SF. That is because the fit has not yet been normalized. Thus, the solution shown in Fig.\ \ref{fig:Dy164_fit_raw} is just one of an infinite set of solutions to the fit. In this section we will discuss how auxiliary data can be used to find the transformation $G$ that gives the physical solutions.

\subsection{Auxiliary experimental data}
\label{sec:Auxiliary_experimental_data}
For the level density $\rho$, there are often two different types of auxiliary datasets available. At low energies, the discrete levels are known from spectroscopy. They can be compared to the fitted level density from the Oslo method after applying the same binning. One can also account for the detector resolution by applying e.g.\  a Gaussian smoothing to the histogram. At higher excitation energies, typically $\sim$1-3~MeV, the spectroscopy data fails to resolve all levels \cite{RIPL3}. The user will thus have to set a sensible region in the low energy regime for the comparison.

The second piece of information stems from neutron resonance experiments, e.g.\ the average $s$-wave resonance spacings $D_0$. They provide information about the level density $\rho(E_x\,{=}\,S_\mathrm{n}, J_\mathrm{t}\pm\onehalf, \pi_\mathrm{t})$ at the neutron separation energy $S_n$, where $J_\mathrm{t
}$ and $\pi_\mathrm{t}$ are the ground-state spin and parity of the target nucleus, i.e.\  the $A-1$ nucleus~\cite{RIPL3}. With the Oslo method we obtain the total level density $\rho(E_x) = \sum_{J, \pi} \rho(E_x, J, \pi)$. If one knows the fraction of $J_\mathrm{t}\pm\onehalf, \pi_\mathrm{t}$ levels, one can estimate $\rho(S_n)$ by from $D_0$ by~\cite{Schiller2000}
\begin{align}
\label{eq:rhoFromD0}
\rho(S_n) = \frac{1}{D_0} \frac{2}{ g(E_x, J_\mathrm{t}+\onehalf) +  g(E_x, J_\mathrm{t}-\onehalf)},
\end{align}
where $g(E_x, J, \pi)$ is the spin-parity distribution of the nucleus at $E_x$ and the factor of 2 comes from the assumption of equiparity, i.e.\  $g(E_x, J, \pi) = g(E_x, J)/2$. Note that the $J_\mathrm{t}-\onehalf$ term vanishes for $J_\mathrm{t}=0$. For the spin-parity distribution $g(E_\mathrm{x},I)$ it is common to use a form proposed in Ref.\ \cite[Eq.\ (3.29)]{Ericson1960}; however the exact parametrization is a major source of systematic uncertainties in the normalization, thus several suggestions are implemented in {\ttfamily OMpy}.

The usage of $\rho(S_n)$ for the normalization is further complicated by restricted fit-regions for $P(E_x, E_\gamma)$ (see Sec.\ \ref{sec:decomposition_fit_range}). These limit the extraction of the level density $\rho$ up to $E_{x}^\mathrm{max}$--$E_{\gamma}^\mathrm{min}$, which is often about 1-3~MeV below $S_n$. Consequently, we cannot directly normalize the fitted level density $\rho$ to $\rho(Sn)$ obtained from $D_0$. To utilize this information, we have to extrapolate $\rho$ and compare the extrapolation at $S_n$. The exact form of the extrapolation is another systematic uncertainty. Generally, the constant temperature (CT) model \cite{Ericson1960} fits well with the level density data obtained from the Oslo method~\cite{Guttormsen2015},
\begin{align}
\rho_\mathrm{CT}(E_x) = \frac{1}{T} \exp \frac{E_x - E_0}{T},
\label{eq:constTemp}
\end{align}
where the temperature $T$ and the energy shift $E_0$ are free parameters of the model. We obtain them by a fit to $\rho$ in a suitable energy range.

The convolution of the level density $\rho$ and $\gamma$-ray strength function $f$ (or equivalently $\mathcal{T}$) can be further constrained by the average total radiative width $\langle \Gamma_\gamma (S_n)\rangle$ from neutron-capture experiments (see e.g. Ref.\ \cite{RIPL3}) on a target nucleus with ground-state spin $J_t$ using the following equation:
\begin{align}
  \label{eq:GammaGamma}
   \langle \Gamma_{\gamma \ell} (S_n) \rangle & = \quad \frac{D_l}{2} \int_0^{S_n} dE_\gamma\, \Biggl[ f(E_\gamma)E_\gamma^3 \rho(S_n-E_\gamma) \notag\\
   & \times \sum_{J_i} \sum_{J_f=|J_i-1|}^{J_i+1} g(S_n-E_\gamma, J_f) \Biggr],
\end{align}
where the first sum runs over all possible residual nucleus spins $J_i$, i.e.\  from $\min|J_t \pm \onehalf \pm \ell|$ to $J_t + \onehalf + \ell$ and the second sum runs over all final spins $J_f$ accessible with dipole radiation starting from a given $J_i$. Often only s-wave information is available, corresponding to $\ell=0$ in the notation above, and the measurements are performed with low energy neutrons, such that $E_x \approx S_n$; for brevity it is then common to write $\langle \Gamma_\gamma\rangle$ for $\langle \Gamma_{\gamma \ell}(S_n)\rangle$. A derivation of Eq.\ \ref{eq:GammaGamma} is given in \ref{app:GammaGamma}. Currently, we approximate $\langle \Gamma_\gamma\rangle$ using the integral involving the level density $\rho$ for all excitation energies $E_x$, although the integral at low energies can be replaced by a sum over decays to the known discrete levels for more precise calculations. Note that we also have to extrapolate the $\gamma$-ray strength function to be able to use this equation. Due to its shape, a log-linear function is often fitted to the results.

\subsection{Likelihood}
In the following, we will define the likelihood $\mathcal{L}(\bm\theta)$ that is used to find the proper normalizations. Let $\mathcal{L}_i(\bm\theta)$ denote the likelihood for a given solution ($\rho$,~$\mathcal{T}$) of Eq.\ \eqref{eq:Pfit_N_rho_T} to match the normalization information~$i$ (i.e.\  discrete levels, $D_0$, $\langle \Gamma_\gamma \rangle$) after the transformation~$G$ with the parameters $\bm\theta=(A, B, \alpha)$. Due to the extrapolation of the level density $\rho$ mentioned above, we have to introduce two nuisance parameters $T$ and $E_0$, so we extend $\bm\theta$ to include $(A, B, \alpha, T, E_0)$. To reduce the computational complexity, we extrapolate the  $\gamma$-ray strength function by the best-fit values for a given set of transformations $(\alpha, B)$. The total likelihood $\mathcal{L}(\bm\theta)$ is then given by
\begin{align}
	\label{eq:likelihood_product}
    \mathcal{L}(\bm\theta) = \prod_i \mathcal{L}_i(\bm\theta).
\end{align}
We assume that the experimental $D_0$ and $\langle \Gamma_\gamma \rangle$ data are normal distributed, thus maximizing the log-likelihood is equivalent to minimizing a sum of $\chi^2_i$'s. Note that the measurement of the discrete levels is of course not stochastic; however, the count data we use to determine $\rho$ for the comparison is. More explicitly, we have
\begin{align}
\label{eq:lnlike_from_chi2}
    \ln \mathcal{L}_i(\bm\theta) &= K_i - \frac{1}{2} \sum_i \chi^2_i, \\
\label{eq:chi2_discrete}
    \chi^2_\mathrm{discrete} &= \sum_j \left(\frac{\rho_{j,\mathrm{discrete}} - \rho_{j,\mathrm{Oslo}}(\bm\theta)}{\sigma_j}\right)^2, \\
\label{eq:chi2_D0}
    \chi^2_{D_0} &= \left(\frac{D_{0,\mathrm{exp}} - D_{0,\mathrm{CT}}(\bm\theta)}
                              {\sigma_{D_{0,\mathrm{exp}}}} \right)^2 \notag \\
                 & \quad + \sum_j \left(\frac{\rho_{j,\mathrm{CT}} - \rho_{j,\mathrm{Oslo}}(\bm\theta)}{\sigma_j}\right)^2, \\
\label{eq:chi2_Gg}
    \chi^2_{\langle \Gamma_\gamma \rangle} &= \left(\frac{\langle \Gamma_\gamma \rangle_\mathrm{exp} - \langle \Gamma_\gamma \rangle_\mathrm{Oslo}(\bm\theta)}
                              {\sigma_{\langle \Gamma_\gamma \rangle_\mathrm{exp}}} \right)^2,
\end{align}
where $K_i=\ln(1/(2\pi\sigma_i))$ is a constant as long as the standard deviation(s) $\sigma_i$ does not depend on $\bm \theta$. The subscript \textit{exp} denotes the experimental data. The sums in Eq.\ \eqref{eq:chi2_discrete} and \eqref{eq:chi2_D0} run over all data points used in the evaluation, and for $\chi^2_{D_0}$ we invert Eq.\ \eqref{eq:rhoFromD0} to obtain $D_0$ from the level density (extrapolated with the CT model). The second term of Eq.\ \eqref{eq:chi2_D0} arises due to the fit of the nuisance parameters of the CT model ($T$, $E_0$) to the Oslo method data, here labeled $\rho_\mathrm{Oslo}$.

As discussed in Sec.\ \ref{sec:decomposition_uncertainty}, the degeneracy of the solutions $\rho$ and $f$ prevents us from directly inferring their parameter uncertainties in the fit of the first generation matrix $P_\mathrm{exp}$. However, clearly the data points of $\rho$ and $\mathcal{T}$ have a statistical uncertainty that should propagate to inform the posterior distribution of $\bm\theta$. We choose to model this by setting  $\sigma_j$ somewhat arbitrarily to a relative uncertainty of 30\%. We propose to test the implications in a future work e.g.\ by comparing inferred $D_0$'s for datasets where $D_0$ is known but on purpose not included in the $\chi^2$ fit of Eq.\ \eqref{eq:chi2_D0}. Note that with this specification of the standard deviations $\sigma_j$, the $K_i$ become $\bm\theta$-dependent:
\begin{align}
\label{eq:Ki_theta}
    K_i = \sum_j \ln \frac{1}{\sqrt{2\pi\sigma_j}} = \sum_j \ln \frac{1}{
    \sqrt{2\pi \times 0.3 \rho_{j,\mathrm{Oslo}}
    (\bm \theta)}
    }.
\end{align}
The likelihood can easily be extended if other information shall be taken into account. In several recent works on the Oslo method, experimental data on $\langle \Gamma_\gamma \rangle$ was not available, but roughly estimated from systematics (see e.g.\ Ref.\ \cite{Larsen2016, Kullmann2019}). With the new possibilities of {\ttfamily OMpy} we would instead recommend to constrain $\alpha$ and $B$ by adding a term to the total likelihood that describes the match with other measured strength function data (usually above $S_n$).

\subsection{Implementation}
We sample this likelihood with the Bayesian nested sampling algorithm {\ttfamily MultiNest}~\cite{Feroz2009, Feroz2013} using the {\ttfamily PyMulti\-Nest} module~\cite{Buchner2014}. For a more efficient calculation, we first find an approximate solution (more accurately the maximum-likelihood estimator) $\bm{\hat\theta}$ with the differential evolution minimizer of {\ttfamily SciPy} \cite{Jones2001, Storn1997}. This is by default used to create weakly informative priors for $A$, $B$ and $\alpha$ and $T$. For $A$ and $B$ we use a normal distribution truncated at 0 (negative values of $\rho$ or $f$ are not meaningful), a default mean $\mu$ given by $\bm{\hat\theta}$ and a broad width of $10\mu$. For $\alpha$ and the nuisance parameter $T$ (entering through the level density extrapolation model) we use log-uniform priors spanning one order of magnitude around $\bm{\hat\theta}$. For the second nuisance parameter $E_0$ we choose a normal distribution with mean 0 and width 5~MeV, which is truncated below (above) -5 and 5~MeV, respectively. The latter choice is well justified regarding the range of reported values of $E_0$ in Ref.\ \cite{Egidy2005}.

This simultaneous normalization is implemented in the {\ttfamily NormalizerSimultan} class, which relies on composition of the {\ttfamily NormalizerNLD} and {\ttfamily NormalizerGSF} classes handling the normalization of the level density $\rho$ and $\gamma$-ray strength function $f$. To facilitate a comparison with the {\ttfamily oslo\hyp{}method\hyp{}software} calculations, one can also run a sequential normalization first using  Eq.\ \eqref{eq:chi2_discrete} and \eqref{eq:chi2_D0} through {\ttfamily NormalizerNLD} and then the resulting $\rho$ as input to the normalization through {\ttfamily NormalizerGSF}, see Eq.\ \eqref{eq:GammaGamma}. It should be stressed though that the normalization through the likelihood calculations in {\ttfamily OMpy} will still differ from the approach taken in the {\ttfamily oslo\hyp{}method\hyp{}software}. The latter allowed only to receive best-fit estimates of the transformation parameters $A$, $B$ and $\alpha$. Any subsequent uncertainty estimation due to the normalization itself was up to the users. We also note that an advantage of the simultaneous approach is that one obtains the correlations between $A$, $B$ and $\alpha$ such that uncertainties in the normalization of $\rho$ directly propagate to the estimation transformation parameters for $f$.

From the {\ttfamily MultiNest} fit we obtain the posterior probability distribution for the parameters $\bm\theta$, given as (equally-weighted) samples $\bm\theta_i$.\footnote{For brevity, we drop the index $m$ on $\bm\theta$, but the normalization is performed for each realization $m$ individually.} The normalization uncertainty for the solution ($\rho^{(m)}$ , $f^{(m)}$) of the realization $m$ can then be mapped out by creating a normalized sample $(\rho^{(m)} , f^{(m)})_i$ for each $\bm\theta_i$ using Eq.\ \eqref{eq:G}. By repeating this procedure for all realizations $m$ of the ensemble, we also recover the uncertainty due to the counting statistics.
%\footnote{In certain situations one may be interested in distinguishing between the uncertainty due to the counting statistics and due to normalization. The former can be obtained if each realization $m$ is transformed by the maximum \text{a posteriori} (MAP) of $\bm\theta$ instead of the different samples $\bm\theta_i$.}
% This does not work!
This is performed in the {\ttfamily EnsembleNormalizer} class.

\section{Systematic uncertainties}
\label{sec:systematic_uncertainties}
The previous sections described the necessary tools to evaluate the statistical uncertainties due to the counting statistics and the normalization procedure. It is important to keep in mind that there are also systematic uncertainties linked to the analysis, which are summarized below:
\begin{enumerate}[a)]
	\item \label{enum:sysunc-removeBg} \textit{Removal of negative counts in raw matrix}: When subtracting the background from the raw matrix, we often receive a matrix with negative counts in some bins. This is clearly linked to the Poisson statistics in regions with a low signal to background ratio. If one simply removes the negative counts, one potentially biases the mean of the level density and $\gamma$-ray strength points derived from these bins. It was previously observed that negative counts in the raw matrix can cause technical challenges in the currently implemented unfolding method, with some bins obtaining extreme negative values after several iterations. For the background subtraction, it might be a more reasonable fix to redistribute the negative counts to bins within the resolution. This is implemented by {\ttfamily Matrix.fill\_negative} as an alternative to the default method, {\ttfamily Matrix.remove\_\allowbreak{}negative}, which removes all negative counts. For the $^{164}$Dy, there is a high signal to background ratio, thus the background subtraction does not have this problem. In cases with a worse background ratio, both methods could be compared. If they result in significant differences, further analysis is needed to find the optimal procedure.

	\item \label{enum:sysunc-removeUnfFg} \textit{Removal of negative counts in unfolded and first\hyp{}generation matrix}: The unfolding and first\hyp{}gen\-er\-a\-tion methods can result in negative counts which can not be linked to the Poisson statistics any longer. It is thus not clear whether it is better to keep the negative counts or to redistribute them in the fashion described above. Again, the problem was negligible in the $^{164}$Dy dataset, but has to be treated carefully if the methods lead to more bins with such a behavior. To retrieve the matrices before removal of the negative counts, the user can simply replace or deactivate the {\ttfamily remove\_\allowbreak{}negative} methods of the {\ttfamily Unfolder} and {\ttfamily FirstGeneration} classes.

	\item \label{enum:sysunc-unfolding} \textit{Unfolding method}: There are two main sources of systematic uncertainties, the iterative unfolding method itself, and the detector response functions. The latter can be gauged by obtaining an ensemble of different detector response functions that capture the breadth of physically reasonable configurations (e.g.\ auxiliary software such as {\ttfamily GEANT4} \cite{Agostinelli2003}). Each member of the raw matrices $R^{(m)}$ is then unfolded with one (or each) of the different detector response functions. The former uncertainty is more difficult to quantify. In this special case, alternative methods exist already and one approach could be to implement an alternative unfolding algorithm (e.g.\ Ref.\ \cite{Choudalakis2012}).
	The {\ttfamily oslo\hyp{}method\hyp{}software} attempts to quantify the systematic uncertainty from unfolding and the first\hyp{}gen\-er\-a\-tion matrix following an \textit{ad hoc} numerical procedure described in Ref.\ \cite{Schiller2000}. A proper treatment that pays tribute to the full complexity of the problem is outside the scope of the present work.

	\item \label{enum:sysunc-PopulationXs} \textit{Population cross-section}: As mentioned in Sec.\ \ref{sec:oslo-method}, the first-generation method assumes that the spin-parity distribution of the populated levels $g_\mathrm{pop}$ is similar for the whole excitation energy range studied. At best, $g_\mathrm{pop}$ approximates the spin-parity distribution $g$ of the levels in the nucleus itself. This is often believed to be the case for low and mid-mass nuclei, where the beam energy is chosen such that the compound cross-section dominates over the direct cross-section. More details can be found in Refs.\ \cite{Larsen2011, Zeiser2019} for challenges if the intrinsic spin-parity distribution $g(E_x, J, \pi)$ is very different from the (normalized) population cross-section $g_\mathrm{pop}(E_x, J, \pi)$.

	\item \label{enum:sysunc-BrinkAxel} \textit{Decomposition and the Brink-Axel hypothesis}: The decomposition of the first-generation matrix into the level density $\rho(E_x)$ and $\gamma$-ray strength function $f(E_\gamma)$ relies on a generalization of the Brink-Axel hypothesis \cite{Brink1955, Axel1962}, where the strength function is assumed to be approximately independent of $E_x$, $J$ and $\pi$ (see Eq.\ \eqref{eq:gsf_Brink_Axel}). The validity of the assumption has been tested within the Oslo method by comparison of the $\gamma$-ray strength function $f$ extracted from different initial and final excitation energy bins, see Refs.\ \cite{Guttormsen2016,Campo2018} and references therein for further works.

	\item \label{enum:sysunc-IntrinsicSpinPar} \textit{Intrinsic spin-parity distribution}: Both the normalization of the level density $\rho$ at $S_n$ via $D_0$ and the absolute scaling of the $\gamma$-ray strength function via $\langle \Gamma_\gamma \rangle$ rely on the knowledge of the intrinsic spin-parity distribution $g(E_x, J, \pi)$ of the nucleus, see Eq.\ \eqref{eq:rhoFromD0} and Eq.\ \eqref{eq:GammaGamma}. It is difficult to measure the spin-parity distribution $g(E_x, J, \pi)$ in the (quasi)continuum and there are various different empirical parametrization and theoretical calculations, see e.g.\ Refs.\ \cite{Egidy2005,Egidy2009,Gilbert1965,Egidy1988} and Refs.\ \cite{Goriely2008, Alhassid2008,Senkov2016,Alhassid2000} respectively.

	\item \label{enum:sysunc-D0Gg} \textit{Lack of $D_0$ and $\langle \Gamma_\gamma \rangle$}: In several recent cases where the Oslo method has been applied, experimental values of $D_0$ and/or $\langle \Gamma_\gamma \rangle$ were not available, see e.g.\ Refs.\ \cite{Larsen2016,Kullmann2019,Renstrom2018,Larsen2017}. In these works, $D_0$ and/or $\langle \Gamma_\gamma \rangle$ were estimated from the values of nearly nuclei. With {\ttfamily OMpy} it is now easier to adopt the normalization procedure via the likelihood $\mathcal{L}(\bm\theta)$ in Eq.\ \eqref{eq:likelihood_product} instead. One can either remove terms where the experimental information is missing, allowing for a larger degeneracy of the solutions, or add different constrains, like a measure of how well the $\gamma$-ray strength function $f(E_\gamma$) matches other measured $\gamma$-ray strength function data, which often exists above $S_n$.

	\item \label{enum:sysunc-BinSizeFitLim} \textit{Bin sizes and fit ranges}: Several decisions have to be made on the bin sizes and fit ranges, for example in the normalization of the level density $\rho$ to the known level scheme, see Sec.\ \ref{sec:Auxiliary_experimental_data}. The user can test different sensible regions for this comparison.
\end{enumerate}

In this work, only the impact of the negative counts, items \ref{enum:sysunc-removeBg} and \ref{enum:sysunc-removeUnfFg} have been studied. In many cases it is challenging to evaluate the impact of the assumptions on the data. Whenever the uncertainty is linked to models, algorithms, parameter sets or alike, and alternative choices exists (items \ref{enum:sysunc-unfolding}, \ref{enum:sysunc-PopulationXs}, \ref{enum:sysunc-IntrinsicSpinPar}, \ref{enum:sysunc-BinSizeFitLim}), it is possible to utilize {\ttfamily OMpy}'s error propagation functionality. This was illustrated for the unfolding of the data in item \ref{enum:sysunc-unfolding}. More information on items \ref{enum:sysunc-BrinkAxel} and \ref{enum:sysunc-IntrinsicSpinPar} can be found in Ref.\ \cite{Larsen2011}.

\section{Discussion and comparison}

\begin{figure*}[bt]
\begin{center}
\includegraphics[clip,width=2.\columnwidth]{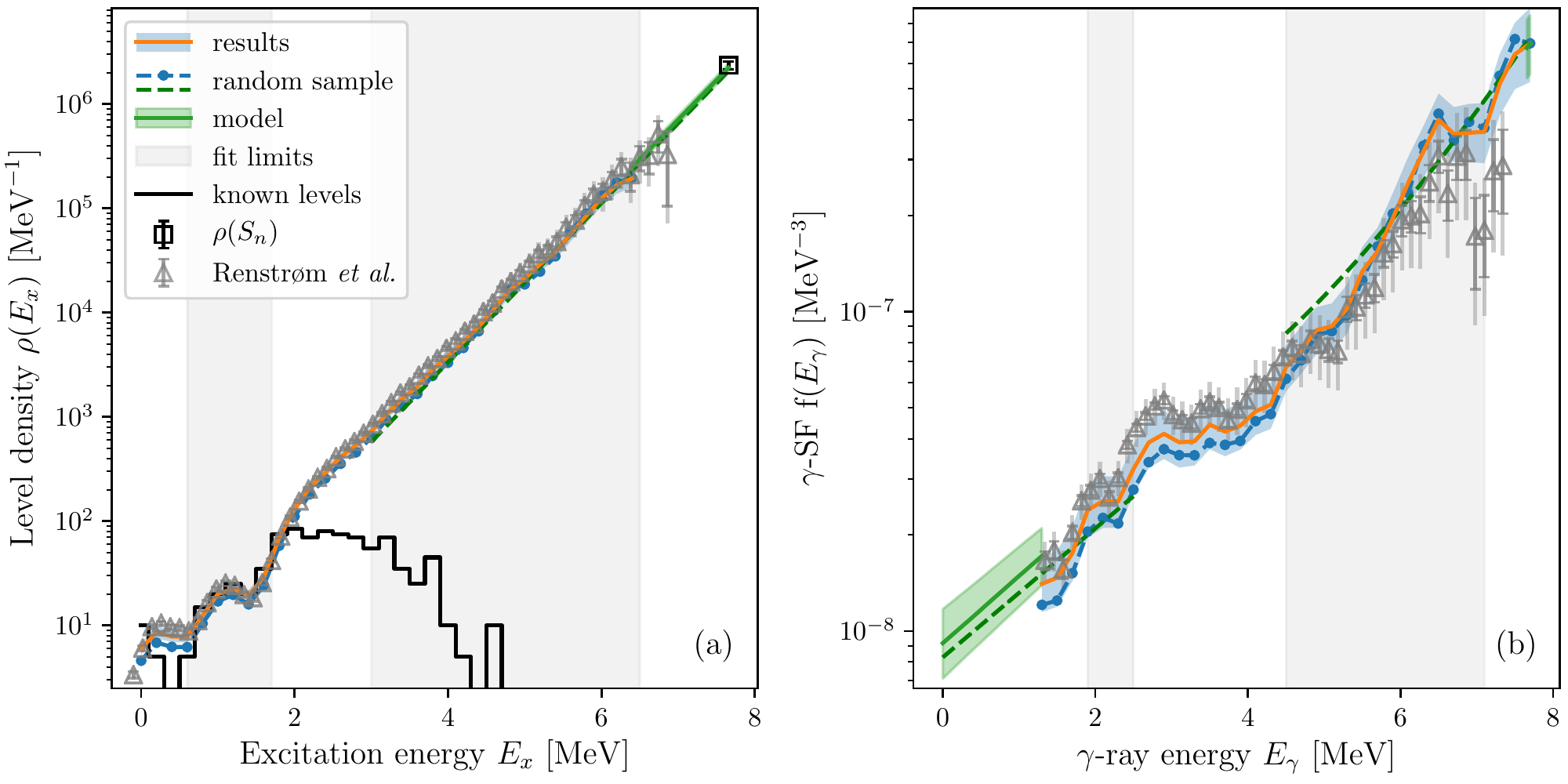}
%\vskip 2cm
\caption{Extracted level density $\rho$ (a) and $\gamma$SF $f$ (b) for $^{164}$Dy. The fit is similar to that shown in Fig.\ \ref{fig:Dy164_fit_raw}, but the normalization according to Eq.\ \eqref{eq:G} has been applied. We display the median and 68\% credible interval obtained from the counting and normalization uncertainties (orange line and blue shaded band, respectively) together with one randomly selected sample of $\rho$ and $f$ (blue dots). The median and 68\% credible interval for the extrapolation is given by the green line and shaded band. In addition, the extrapolation used together with the random sample is indicated by the dashed line. The data points within the gray area denoted \textit{fit limits} are used for the normalization and extrapolations, such that we match the binned known levels (black line) \cite{NNDC}, $\rho(S_n)$ calculated from $D_0$ \cite{RIPL3} (and $\langle \Gamma_\gamma \rangle$ \cite{RIPL3}). The results are compared to the analysis of \citet{Renstrom2018} which used the {\ttfamily oslo\hyp{}method\hyp{}software}, displaying both the uncertainty that is quoted due to the count statistics (inner error bar) and the total uncertainty, including the normalization (outer error bar) (see text for more details).}
\label{fig:Dy164_fit}
\end{center}
\end{figure*}

In Fig.\ \ref{fig:Dy164_fit} we show the level density $\rho$ and $\gamma$-ray strength function $f$ of $^{164}$Dy resulting from the simultaneous normalization of  $N_\mathrm{ens}=50$ realizations. Each realization $m$ is transformed with $N_\mathrm{samp}=100$ samples from the normalization parameters $\bm\theta_i$. The combined uncertainty of the normalization and counting statistics is visualized through the 16th, 50th, and 84th percentiles which together form the median and a 68\% credible interval. Additionally, we show one randomly selected sample $(\rho^{(m)} , f^{(m)})_i$ including its extrapolation. The results are compared to the analysis of \citet{Renstrom2018} which used the {\ttfamily oslo\hyp{}method\hyp{}software}, and we display both the uncertainty that is quoted due to the counting statistics and the total uncertainty, that includes the normalization. Note that we discussed in Sec.\ \ref{sec:decomposition_uncertainty} that this split into the uncertainty due to counting statistics and due to normalization is, strictly speaking, not possible -- hence, the quoted decomposition is an approximation.

It is gratifying to see that overall, both analyses provide similar results. In Fig.\ \ref{fig:Dy164_fit}a, the level density below $\sim 2$\,MeV exhibits the same structure of bumps attributable to the discrete level structure, and at higher $E_x$ the curves are practically identical. Similarly, for the $\gamma$SF in Fig.\ \ref{fig:Dy164_fit}b, the data is mostly compatible within the error-bars. The median of the results from {\ttfamily OMpy} has slightly steeper slope and the uncertainties are somewhat more evenly distributed across the whole energy range.

Some deviation between the results is expected due to the different fitting method and software implementations. Moreover, different fit-regions may lead to different estimations of the normalization parameters $\theta$, which in turn give different slopes (and absolute values) for $\rho$ and the $\gamma$SF. Usually, there will be several sensible fit-regions. With {\ttfamily OMpy}, the user could create a wrapper that loops through the different fit-regions. The mean and spread of the results can be analyzed using the same ensemble based approach as above.

Another way to verify the results of {\ttfamily OMpy} is to simulate decay data for a given $\rho$ and the $\gamma$SF and compare these to the analysis with {\ttfamily OMpy} from the simulated data. We have used the Monte-Carlo nuclear decay code {\ttfamily RAINIER} v1.5~\cite{Kirsch2018, RAINIER_v1.5} to create decay data from a  $^{164}$Dy-like nucleus. To create an artificial level scheme, we used the  first 20 known discrete levels \cite{RIPL3} and for higher energies created levels following the CT model and spin-distribution described in \citet{Renstrom2018}. The $\gamma$SF has been modeled with the parameters from the same publication. Then, we simulate the experiment by populating $2\times10^6$ levels below $S_n$ and recording the decay $\gamma$s for each event. For simplicity, we have assumed here that we populate the levels proportionally to the intrinsic spin distribution of the nucleus. Finally, we use the response functions of the $^{164}$Dy experiment to convert the recorded $\gamma$ rays to a matrix of synthetically generated events that substitute the experimentally determined raw matrix in the further analysis with {\ttfamily OMpy}. The full setting file can be found with the supplementary material \href{https://doi.org/10.24433/CO.6094094.v1}{online}.

\begin{figure*}[bt]
\begin{center}
\includegraphics[clip,width=2.\columnwidth]{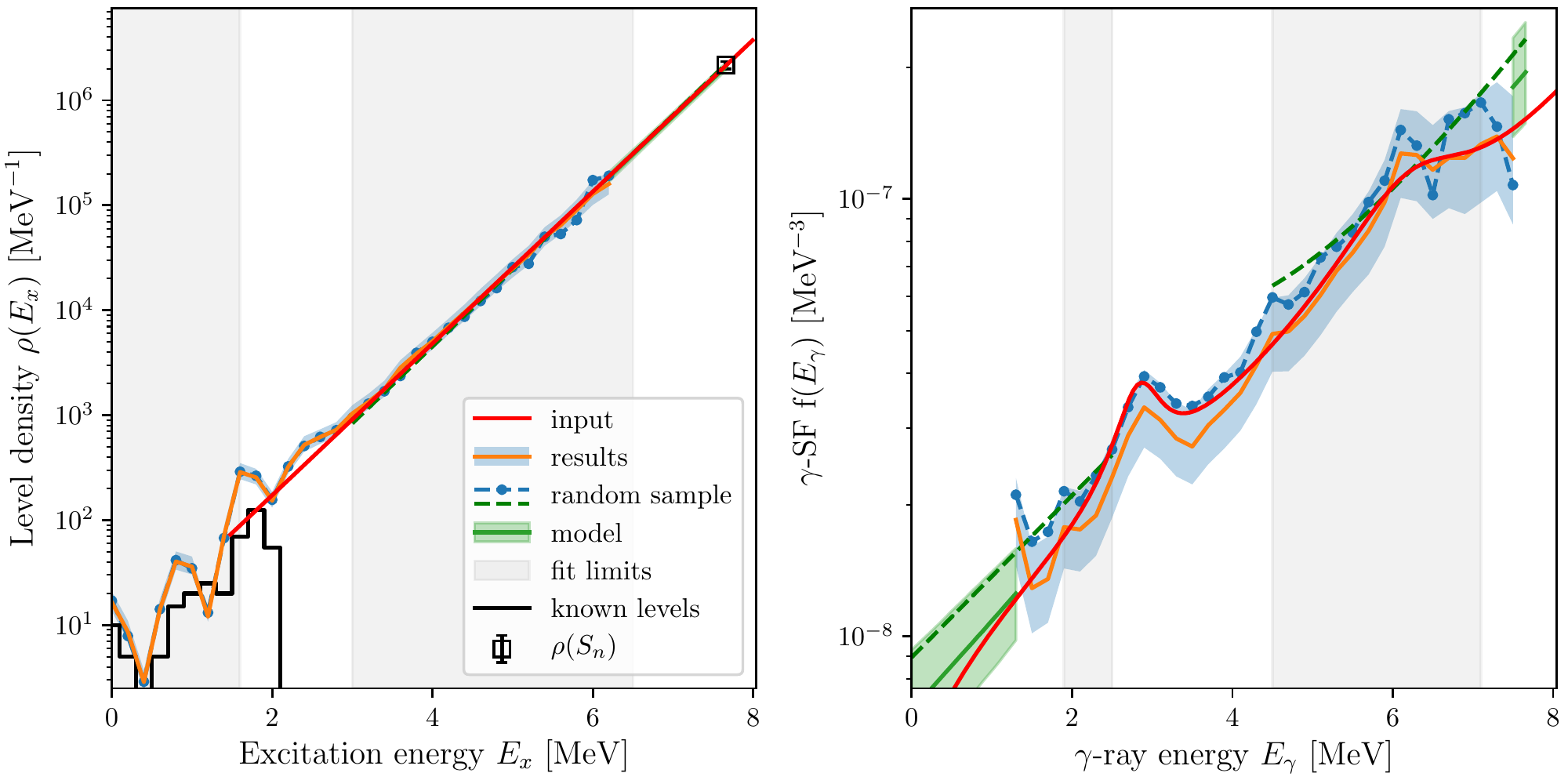}
%\vskip 2cm
\caption{Extracted level density $\rho$ (a) and $\gamma$SF $f$ (b) from synthetic data for a $^{164}$Dy-like nucleus. In the quasicontinuum region the analysis with {\ttfamily OMpy} (labeled results) agrees well with the input level density and $\gamma$-ray strength function model.}
\label{fig:Dy164_synthetic}
\end{center}
\end{figure*}

In Figure \ref{fig:Dy164_synthetic} we compare the results of {\ttfamily OMpy} to the input level density $\rho$ and $\gamma$SF from RAINIER. The fitted level density $\rho$ slightly over-pronounces the structures of the discrete levels at low excitation energies. From a practical point of view this is not a significant problem, as the Oslo method is used for an analysis of the quasicontinuum region, so at energies $\gtrsim 2$ MeV. There, we observe a perfect match between the input and the fitted level density. The resulting $\gamma$SF is in very good agreement with the input $\gamma$SF. There are small deviations at the lowest and highest energies that can easily be understood. The apparent discrepancy of the input  $\gamma$SF below $\sim 2$~MeV can be explained by a failure of the first-generation method due to strong populations of discrete states and is directly visible in the comparison of the unfolded to first-generation matrix (see supplementary material \href{https://doi.org/10.24433/CO.6094094.v1}{online}). The region could have been excluded from the comparison, but we aimed for the same extraction region as for the experimental $^{164}$Dy data set. The mismatch propagates to a wrong extrapolation of the $\gamma$SF towards lower energies. The two resonance-like structures at $\sim$6 and 7 MeV over-compensate for the mismatch of the level density at the lowest excitation energies.

\begin{figure*}[bt]
\begin{center}
\includegraphics[clip,width=2.\columnwidth]{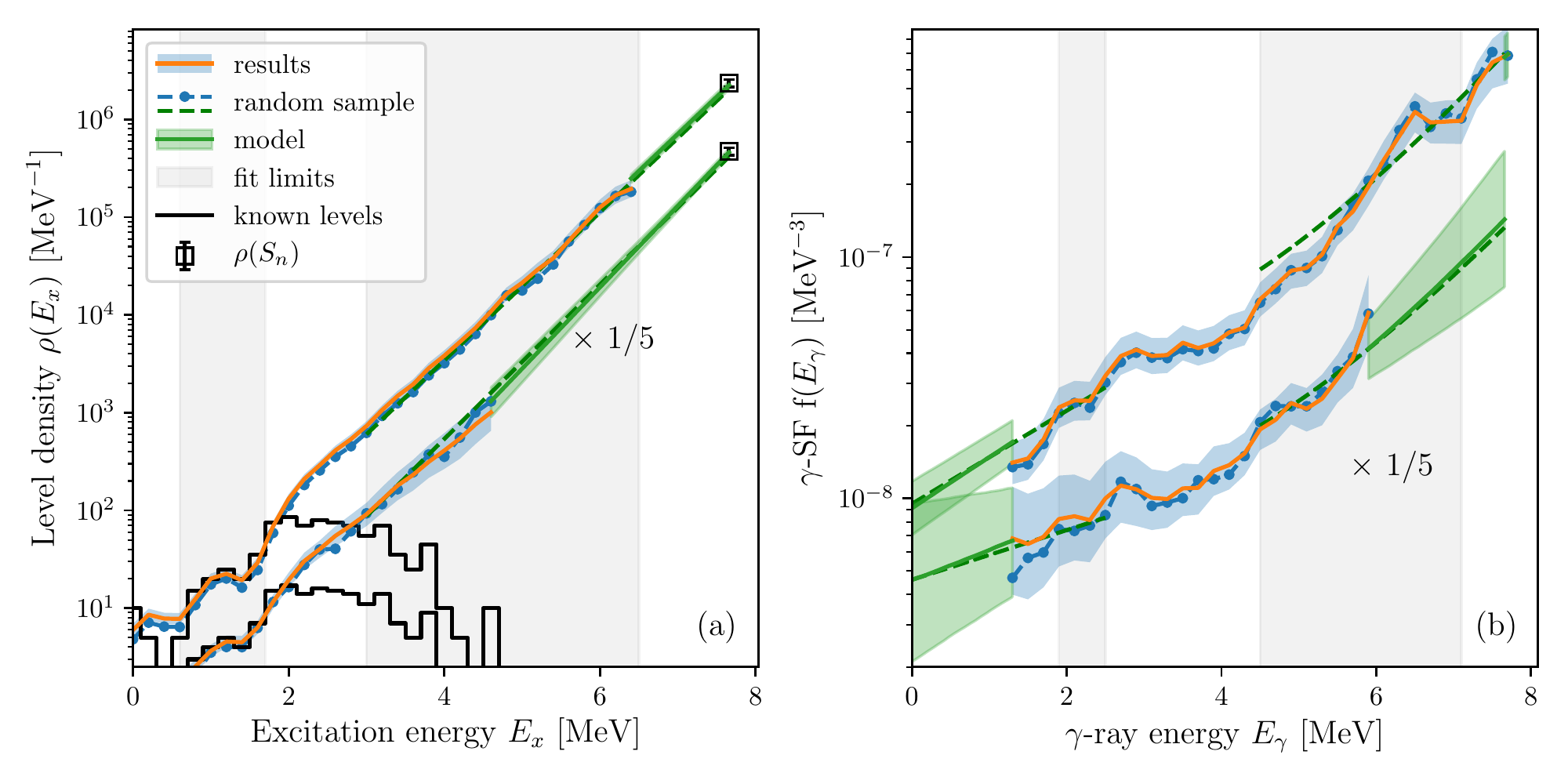}
%\vskip 2cm
\caption{Similar to Fig.\ \ref{fig:Dy164_fit} we show the extracted level density $\rho$ (a) and $\gamma$SF $f$ (b) for the $^{164}$Dy experiment, but here it is compared to a case with only 1/10th of the data. The level density $\rho$ (a) and $\gamma$SF from the case with reduced number of counts have been scaled down by a factor of 5 in the plot to facilitate the visual comparison with the original analysis. For each case, one randomly selected sample including its extrapolation is shown in addition to the median and a 68\% credible interval.}
\label{fig:Dy164_comparison}
\end{center}
\end{figure*}

One of the main motivations for {\ttfamily OMpy} was to improve the uncertainty analysis in the Oslo method. Throughout the article we have highlighted several improvements to the theoretical framework for propagating the uncertainties compared to the {\ttfamily oslo\hyp{}method\hyp{}software}. The most fundamental source of uncertainties in the analysis is the count statistics. For each experiment a balance between the run-time costs and the possibility to gather more data has to be found. Therefore, we find it instructive to study the impact of a reduced number of counts for the $^{164}$Dy experiment. We create a new raw matrix, which we draw from a Poisson distribution with a mean of 1/10th of the original data (an arbitrary choice corresponding to 1/10th of the run actual time) and otherwise perform the same analysis as above. Both analyses are compared in Fig.\ \ref{fig:Dy164_comparison}, and overall we find a good agreement of the median values.

There is one major difference in the analysis that leads to different uncertainty estimates. In the case with the reduced number of counts we have to use a lower  maximum excitation energy $E_x^\mathrm{max}$ in the fit to $P(E_x, E_\gamma)$. As a  consequence, the extracted level density ranges only up to about 4.5 MeV instead of 6.5 MeV, increasing the uncertainty in the determination of the normalization parameters. The increased uncertainty of the slope parameter $\alpha$ is directly visible for the level density $\rho$. For the $\gamma$SF we additionally require a suitable fit to the experimental $\langle \Gamma_\gamma \rangle$, so a different slope $\alpha$ will also affect the absolute scaling $B$. This results in the normalized $\gamma$SFs \textit{tilting} around a $\gamma$-ray energy of about 5 MeV, where the exact location of the tilting point depends on the nucleus. The normalization uncertainty thus leads to increasing relative uncertainties of the $\gamma$SF for lower $\gamma$-ray energies $E_\gamma$. This contrasts to the naive expectation that the relative uncertainties should increase with $E_\gamma$, as the number of counts that determine each $\gamma$SF bin decrease. Finally, we foresee that the updated framework for the uncertainty analysis may have a significant impact when further processing of the results from the Oslo method. An example of this is given in \citet{Renstrom2018}, where the authors fit of the strength of the peak at $\sim3$~MeV in the $\gamma$SF which assumed to be a $M1$ scissors mode.

\section{Extension of {\ttfamily OMpy}}
{\ttfamily OMpy} is written with modularity in mind. We want it to be as easy as possible for the user to add custom functionality and interface {\ttfamily OMpy} with other libraries. For example, in Sec.\ \ref{sec:systematic_uncertainties} we discussed that it may be of interest to try other unfolding algorithms than the one presently implemented. To achieve this, one just has to write a wrapper function with the same return structure as the callable {\ttfamily Unfolder} class. Then one provides the new wrapper to {\ttfamily Ensemble} instead of the {\ttfamily Unfolder} class and all matrices will be unfolded with the new algorithm.

It is our hope and goal that {\ttfamily OMpy} will be used as much as possible. Feedback and suggestions are very welcome. We encourage users who implement new features to share them by opening a pull request in the \href{https://github.com/oslocyclotronlab/ompy}{Github repository}.

%%%%%%%%%%%%%%%%%%%%%%%%%%%%%%
% Conclusions and outlook
%%%%%%%%%%%%%%%%%%%%%%%%%%%%%%

\section{Conclusions and outlook}
We have presented {\ttfamily OMpy}, a complete reimplementation of the Oslo method in Python. The capabilities of the code have been demonstrated by comparison with a synthetic data set modeled with the decay code {\ttfamily RAINIER}. We have also compared {\ttfamily OMpy} to the analysis for $^{164}$Dy with the previous implementation of the Oslo method, the {\ttfamily oslo\hyp{}method\hyp{}software} and in general find a good agreement. However, we have refined the uncertainty quantification of the analysis using an ensemble-based approach. We are now able to simultaneously take into account statistical uncertainties from the counting statistics and the normalization to external data and preserve the full correlation between the resulting level density $\rho$ and $\gamma$SF.

\section*{Acknowledgements}
We would like to thank V. W. Ingeberg for a beta test and his feedback on the software. This work was supported by the Research Council of Norway under project Grants No. 263030, 262952 and 23054. A.C.L. gratefully acknowledges funding from the European Research Council, ERC-STG-2014 Grant Agreement No. 637686.

%\clearpage
\appendix
\section{Unfolding}
\label{app:unfolding}
Here we explain the unfolding technique presented in Ref.\ \cite{Guttormsen1996}, which is used both in the original Oslo method implementation in the {\ttfamily oslo\hyp{}method\hyp{}software} and in {\ttfamily OMpy}.
Let the detector response be modeled as a conditional probability density function
\begin{align}
    p(E_\gamma | E_\gamma'),
\end{align}
encoding the probability that a $\gamma$ ray with true energy $E_\gamma'$ is detected with energy $E_\gamma$. Given a true $\gamma$-ray spectrum $U(E_\gamma)$, the folded spectrum $F(E_\gamma)$, i.e.\ the spectrum seen by the detector, is then given by
\begin{align}
    F(E_\gamma) = \int p(E_\gamma | E_\gamma') U(E_\gamma') \,\mathrm{d}E_\gamma'.
\end{align}
By discretizing into energy bins of width $\Delta E_\gamma$, it becomes a matrix equation
\begin{align}
    \vec{F} = P \vec{U},
\end{align}
where $P$ is the response matrix of discrete probabilities $P_{kl} = p(E_{\gamma,k}|E_{\gamma,l}') \Delta E_\gamma$.
The unfolding procedure amounts to solving this equation for $\vec U$. However, a straightforward matrix inversion is ill-advised, as it will often lead to singularities or produce large, artificial fluctuations in $\vec U$ \cite{Kaipio2007, Blobel2013}. Instead, the approach taken in the Oslo method is to use an iterative technique that successively approximates $\vec U$. Letting $\vec R$ denote the measured spectrum, the algorithm is
\begin{enumerate}
    \item Start with a trial function $\vec U_0 = \vec R$ at iteration $i=0$
    \item Calculate the folded spectrum $\vec F_i = P \vec U_i$ \label{it:2}
    \item Update the trial function to $\vec U_{i+1} = \vec U_i + (\vec R - \vec F_i)$ \label{enum:unfolding_additive}
    \item Iterate from \ref{it:2} until $\vec F_i \approx \vec R$. \label{enum:unfolding_end}
\end{enumerate}
Note that the {\ttfamily oslo\hyp{}method\hyp{}software} uses a custom tailored combination of the additive updating procedure of step \ref{enum:unfolding_additive}, and a ratio approach, $\vec U_{i+1} = \vec U_i \circ (\vec R \oslash \vec F_i)$, where the Hadamard products stand for an element wise product. We obtain equally good results by adopting only the additive updating in {\ttfamily OMpy}. The updating procedure may sometimes lead to a negative number of counts in the unfolded spectra. For negative counts close to zero, it is not clear whether these should be kept or discarded, as it is not clear whether they originate from the statistical nature of the data. In some cases, one observes some bins with large, negative counts which hints at a failure of the method. These cases should be analyzed carefully before the results are processed further. The current default is to remove the negative counts at the end of the unfolding.

A too large number of iterations does not improve the results significantly, but introduces strong fluctuations in the unfolded spectrum.
After the publication of Ref.\ \cite{Guttormsen1996} a criterion for step \ref{enum:unfolding_end} has been added to the {\ttfamily oslo\hyp{}method\hyp{}software}, which is also used in {\ttfamily OMpy}. A predefined number (usually around 30-200) iterations is run and the best iteration is selected based on a weighted sum over each vector element of the root-mean-square error of $\vec F_i - \vec R$ and the relative level of fluctuations in $\vec U_i$ compared to the fluctuations of the raw spectrum $\vec R$. The relative fluctuations are estimated as $\lvert (\vec U_{i,l} - \langle\vec{U_{i}}\rangle)/\langle\vec{U_{i}}\rangle\rvert_1$, where $\langle\vec{U_{i}}\rangle$ is a smoothed version of the spectrum $\vec U_i$.

In addition to this, the user can choose to use a further refinement to the unfolding method known as Compton subtraction~\cite{Guttormsen1996}. It is used to further control the fluctuations in the unfolded spectrum. The basic concept behind it is to use the previously unfolded spectrum to decompose $\vec R$ into parts corresponding to the full-energy, single and double escape and annihilation peaks, and the ``rest'' which comes from Compton scattering and similar processes. Each of these parts, save for the full-energy peak, are then smoothed with the detector resolution before they are subtracted from $\vec R$. The resulting spectrum normalized to maintain the number of counts. The idea is that this should give an unfolded spectrum with the same statistical fluctuations as in the original spectrum $\vec R$.

\section{The first-generation method}
\label{app:firstgen}

In this appendix we describe the idea behind the first-generation method of Ref.\ \cite{Guttormsen1987} and its implementation in {\ttfamily OMpy}. Let $FG(E_\gamma)_{E_x}$ denote the first-generation $\gamma$-ray spectrum, i.e., the intensity distribution of $\gamma$-ray decay from a given excitation energy $E_x$, as function of $\gamma$-ray energy $E_\gamma$.
Generally, the nucleus will decay from $E_x$ down to the ground state by emitting a cascade of $\gamma$ rays, which forms the \emph{total} $\gamma$-ray spectrum.
The total, or all-generations $\gamma$-ray spectrum, denoted $AG(E_\gamma)_{E_x}$, can be viewed as a superposition of the first-generation spectrum and a weighted sum of the all-generations spectra of excitation energies below,
\begin{align}
    AG(E_\gamma)_{E_x} &= FG(E_\gamma)_{E_x}\\
    &+ \sum_{E_x' < E_x} w(E_x')_{E_x} \underbrace{n(E_x')_{E_x}  AG(E_\gamma)_{E_x'}}_{\langle AG(E_\gamma)_{E_x'}\rangle}.\nonumber
    \label{eq:AG_FG}
\end{align}
Here, $w(E_x')_{E_x}$ is a weight factor that gives the decay probability from $E_x$ to $E_x'$, and $n(E_x')_{E_x}$ is a normalization factor which corrects for the varying cross section to populate the $E_x'$ bins. Note that the normalization factor $n(E_x')_{E_x}$ times the all-generations spectrum $AG(E_\gamma)_{E_x'}$ gives the spectra of $\gamma$ rays $\left\langle AG(E_\gamma)_{E_x'}\right\rangle$ emitted from the $E_x'$ bin for each single population.
\subsection*{Normalization and multiplicity estimation}
The normalization factor $n(E_x')_{E_x}$ can be estimated from the singles spectra $S(E_x)$, which contain the number of reactions populating the excitation energy bin $E_x$ and thus the number of $\gamma$-ray cascades out of this level,
\begin{align}
    n(E_x')_{E_x} = \frac{S(E_x)}{S(E_x')}.
\end{align}
However, usually it is determined from the total $\gamma$-ray spectrum by the relation
\begin{align}
    n(E_x')_{E_x} = \frac{\langle M(E_x')\rangle N(E_x)}{\langle M(E_x)\rangle N(E_x')},
\end{align}
where $\langle M(E_x)\rangle$ and $N(E_x)$ denote the average $\gamma$-ray multiplicity and the total number of counts, respectively, at excitation energy $E_x$. This reformulation uses the fact that $S(E_x) = N(E_x) / M(E_x)$. In {\ttfamily OMpy} there are two ways to determine the average multiplicity $\langle M(E_x)\rangle$. The initial idea, the \textit{total multiplicity estimation} is given in  Ref.\ \cite{Rekstad1983} as
\begin{align}
    \langle M(E_x)\rangle = \frac{E_x}{\langle E_\gamma \rangle},
    \label{eq:fg_total_multifilcity}
\end{align}
where $\langle E_\gamma \rangle$ is the weighted-average $\gamma$-ray energy at excitation energy $E_x$.
Due to the detector threshold, we are usually not able to measure all $\gamma$ rays, and this will artificially increase $\langle E_\gamma \rangle$. To solve this problem, a \textit{statistical multiplicity estimation} has been added to the {\ttfamily oslo\hyp{}method\hyp{}software} and is adapted in {\ttfamily OMpy}. The underlying idea is that in heavier nuclei, like in the rare-earth region, $\gamma$ rays from entry states at higher excitation energy will decay down to the yrast-line, where it enters at an energy denoted here as $E_\mathrm{yrast}$. From there, the $\gamma$ rays follow a non-statistical decay to the ground state. For heavy nuclei, there are many levels at low excitation energies, such that it is assumed here that the yrast transitions will proceed with many $\gamma$ rays of so low energy, that they are usually below the detector threshold. In that case, we can replace the excitation energy in Eq.\ \eqref{eq:fg_total_multifilcity} by the \textit{apparent} excitation energy $\tilde E_x = E_x - E_\mathrm{yrast}$. A challenge in this method is to correctly estimate the entry energy $E_\mathrm{yrast}$. Thus we recommend to use the \textit{total multiplicity estimation} whenever the experimental conditions allow for it.

\subsection*{Weight function and iteration}
The weight function $w(E_x')$ encodes the probability for the nucleus to decay from $E_x$ to $E_x'$, and is in fact nothing but the normalized first-generation spectrum for $E_x$,
\begin{align}
w(E_x')_{E_x} = \frac{FG(E_x - E_x')_{E_x}}{\sum_{E_\gamma'}FG(E_\gamma')_{E_x}}.
\end{align}
By rewriting Eq.\ \eqref{eq:AG_FG}, we obtain
\begin{align}
    &FG(E_\gamma)_{E_x} = AG(E_\gamma)_{E_x}  \label{eq:fg_from_ag}\\
    &- \sum_{E_x' < E_x} n(E_x')_{E_x} \frac{FG(E_x - E_x')_{E_x}}{\sum_{E_\gamma'}FG(E_\gamma')_{E_x}} AG(E_\gamma)_{E_x'}.\nonumber
\end{align}
This is a self-consistent set of equations for the $FG$ spectra, which we solve by an iterative procedure, starting with a set of trial functions $FG(E_\gamma)_{E_x}$ and iterating until convergence is reached. In {\ttfamily OMpy}, the trial functions are chosen as constant functions, i.e.\ ~with the same value for all $E_\gamma$. In the original implementation of the first-generation method, the trial functions are instead chosen based on a Fermi gas level density model \cite{Bethe1937, RIPL3}. We have checked with {\ttfamily OMpy} that this gives identical results as with constant functions. The iterative procedure may sometimes lead to a negative number of counts in the resulting first generation spectra. As for the unfolding method, it is not clear whether they originate from the statistical nature of the data or from inaccuracies in the iterative procedure. The current default solution in {\ttfamily OMpy} is to remove negative bins from the obtained FG spectra. However, as described in Section 7, the user can choose to rather redistribute the negative counts to neighboring bins using the {\ttfamily Matrix.fill\_negative} option. In cases where a few bins obtain large, negative counts, we advise the user to carefully analyze the results before processing them further.

Based on the development in {\ttfamily oslo\hyp{}method\hyp{}software} a small variation was added to {\ttfamily OMpy} to ensure better convergence for higher iteration numbers. Starting from iteration 5, each iteration $i$ of the $FG$ spectra is calculated at $FG_{i} = 0.7{FG}'_i + 0.3 FG_{i-1}$, where ${FG}'_{i}$ is the spectrum from the analytical solution of \eqref{eq:fg_from_ag}. This is comparable to a fixed learning rate in machine learning algorithms.

\section{Derivation of the Oslo method equation}
\label{app:oslomethodeq}
Here, we derive the relationship between the distribution of primary $\gamma$ rays, and the strength function and level density. The derivation is based on \citet[p.\ 214--217]{Weisskopf1943}, and \citet[p.\ 342, 649]{Blatt1952}, but generalizes the equations to take into account angular momentum conservation. This corresponds to the Hauser-Feshbach theory of statistical reactions \cite{Hauser1952}, but we apply several simplifications, like the Brink-Axel hypothesis, such that we can arrive at the \textit{Oslo method equation}, Eq.\ \eqref{eq:Oslo_method_eq}.

We start with Bohr's independence hypothesis \cite{Bohr1936} for the cross section $\sigma$ in a nuclear reaction $a + A \to \mathrm{CN}^* \to c + C$.\footnote{More precisely, we assume that our reaction, e.g.\ a $(d,p)$ reaction, leads to a compound nucleus $\mathrm{CN}^*$, which subsequently decays by $\gamma$-ray emission.} Let us denote all quantum numbers of the entrance channel, the compound nucleus and the exit channel by $\alpha$, $i$ and $f$, respectively. The cross section is then given by \cite[p.\ 342]{Blatt1952}, \cite{Bohr1936}
\begin{align}
    \label{eq:compound xs single J}
    \sigma(\alpha, f) = \sigma_{\mathrm{CN}}(\alpha) G_{\mathrm{CN}}(i, f),
\end{align}
where the compound nucleus formation cross-section $\sigma_{\mathrm{CN}}$ depends only on the entrance channel $\alpha$. Following Bohr, the decay probability of the compound nucleus $G_{\mathrm{CN}}$  depends only on the branching ratios of the compound nucleus level $i$ to a specific channel $f$, but not on the entrance channel. In the Oslo method, we select only excitation energies $E_x$ below the particle separation threshold, so the compound nucleus can decay by $\gamma$-rays only. The decay probability $G_{\mathrm{CN}}$ of the state $i$ is then simply the $\gamma$-ray branching ratio to a specific final state $f$ with energy $E_f$ and the spin-parity $J^\pi_f$,
\begin{align}
    \label{eq:branching ratio}
    G_{\mathrm{CN}} = \frac{\Gamma_\gamma(i \to f)}{\Gamma_\gamma},
\end{align}
where $\Gamma_\gamma(i \to f)$ is a partial, and $\Gamma_\gamma=\sum_f \Gamma_\gamma(i \to f)$ is the total radiative width for the level $i$.

The spectrum of the decay radiation $n(i, f)\mathrm{d}E_f$ from the compound nucleus level $i$ is then given by the summation (or integration) of $\sigma(\alpha, f)$ over an interval $dE_f$ (up to a constant due to the flux of $a$ and density of $A$ which would cancel out later):
\begin{align}
    \label{eq:spectrum beta}
    n(i, f)\mathrm{d}E_f &= \sum_{f \mathrm{~in~} E_f} \sigma(\alpha, f)
                   = \sigma_{\mathrm{CN}}(\alpha) \sum_{f \mathrm{~in~} E_f} \frac{\Gamma_\gamma(i \to f)}{\Gamma_\gamma} \notag \\
    &= \sigma_{\mathrm{CN}}(\alpha) \sum_{XL} \frac{\left\langle \Gamma_\gamma(i \to f) \right\rangle}{\Gamma_\gamma^{(L)}} \rho_\mathrm{avail}(E_f), \notag \\
    &= \sigma_{\mathrm{CN}}(\alpha) \sum_{XL} \frac{f_{XL}(E_\gamma) E_\gamma^{2L+1}}{\Gamma_\gamma^{(L)}\rho(E_x, J^\pi_i)}  \rho_\mathrm{avail}(E_f) \notag \\
    &= C_{\alpha, i} \sum_{XL} \frac{f_{XL}(E_\gamma)}{\Gamma_\gamma^{(L)}} E_\gamma^{2L+1}  \rho_\mathrm{avail}(E_f),
\end{align}
where $\rho_\mathrm{avail}$ is density of accessible final levels at $E_f$, $C_{\alpha \to i}$ is a constant that depends only on the entrance channel (which determines the compound nucleus state $i$), and we have replaced the average partial radiative width $\langle \Gamma_\gamma(i \to f) \rangle$ by $\gamma$-ray strength-function\footnote{To keep standard notation we will denote both the $\gamma$-ray strength function and the final level by $f$. It should be clear from the context what we refer to.} $f_{XL}$.

The $\gamma$-ray strength function $f_{XL}$ for a given multipolarity $XL$ and for decays from an initial level $i$ to final level $f$ is defined as \cite{Bartholomew1973}
\begin{align}
    \label{eq:gsf_definition}
    &f_{XL}(E_x, J^\pi_i, E_\gamma, J^\pi_f) = \frac{\left\langle \Gamma_{\gamma XL}(E_x, J^\pi_i, E_\gamma, J^\pi_f \right\rangle}{E_\gamma^{2L +1}} \rho(E_x, J, \pi)
\end{align}
where $\langle \cdots \rangle$ denotes an average over individual transitions in the vicinity of $E_x$ (in practice defined by the energy binning resolution). This can be simplified using the dominance of dipole radiation ($L=1$) and a common generalization of the Brink-Axel hypothesis \cite{Brink1955, Axel1962}, where the strength function is assumed to be approximately independent of $E_x$, $J$ and $\pi$,
\begin{align}
    \label{eq:gsf_Brink_Axel}
    \sum_{XL} f_{XL}(E_x, J^\pi_i, E_\gamma, J^\pi_f) & \mathrel{\overset{\makebox[0pt]{\mbox{\normalfont\tiny\sffamily dipole}}}{~\approx~}} f_1(E_x, J^\pi_i, E_\gamma, J^\pi_f) \notag \\
    & \mathrel{\overset{\makebox[0pt]{\mbox{\normalfont\tiny\sffamily Brink-Axel}}}{~\approx~}} f_1(E_\gamma)
\end{align}
where we define the total dipole strength function $f_1$ as the sum of the electric and magnetic dipole strength, $f_{E1}$ and $f_{M1}$, respectively, $f_1=f_{E1} + f_{M1}$.

If we again use the dominance of dipole radiation in Eq.\ \eqref{eq:spectrum beta} (which leads to $\Gamma_\gamma \approx \Gamma_\gamma^{(L=1)}$), and assume parity equilibration of the level density, i.e.\ ~$\rho(E_x, J, +) \approx \rho(E_x, J, -)$ we can write
\begin{align}
    \label{eq:spectrum beta simple}
    & n(i, f) \mathrm{d}E_f \approx C_{\alpha, i} E_\gamma^3 \notag \\
    & \quad \times \left(  \frac{f_{E1}(E_\gamma)}{\Gamma_\gamma}  \sum_{J_f=J_i-1}^{J_f=J_i+1} \rho(E_f, J_f, -\pi_i) \notag \right.\\
    & \phantom{\quad \times \big(} + \left.
                \frac{f_{M1}(E_\gamma)}{\Gamma_\gamma} \sum_{J_f=J_i-1}^{J_f=J_i+1} \rho(E_f, J_f, \pi_i) \right) \notag \\
    & \quad = C'_{\alpha, i} E_\gamma^3 f_1(E_\gamma) \sum_{J_f=J_i-1}^{J_f=J_i+1} \rho(E_f, J_f, \mathrm{eq}),
\end{align}
where $\rho(E_x,J_f,\mathrm{eq})$ denotes the level density of one parity, the notation emphasizing the assumption of parity equilibration\footnote{The selection rules dictate that dipole radiation changes the angular momentum $J$ by at most one unit. For $M1$, the parity is unchanged, while for $E1$ it flips. This determines the density of available final levels for the decay. In the case of $J_i=1/2$ the sum runs over $J_f=\{ 1/2, 3/2 \}$, and in the case of $J_i=0$, the sum only runs over $J_f=1$, since $J=0 \to J=0$ transitions are forbidden. \label{fn:selection rules}} and $C'_{\alpha, i}=C_{\alpha, i}/\Gamma_\gamma$. We may write the partial level density $\rho(E_x, J, \mathrm{eq})$ as
\begin{align}
    \label{eq: spin dist eq}
    &\rho(E_x, J, \mathrm{eq}) = \frac{1}{2} g(E_x, J) \rho(E_x),
\end{align}
where $g$ denotes the intrinsic spin distribution of the nucleus and $\rho(E_x)=\sum_{J^\pi}\rho(E_x, J,\pi)$ is the ``summed'' (or ``total'') nuclear level density. With this, we can further simplify the sum over the final levels in Eq.\ \eqref{eq:spectrum beta simple}:
\begin{align}
   \label{eq:sum final states}
   & \sum_{J_f=J_i-1}^{J_f=J_i+1} \rho(E_f, J_f, \mathrm{eq}) \\
   & \quad =  \frac{\rho(E_f)}{2} \sum_{J_f=J_i-1}^{J_f=J_i+1} g(E_f, J_f) \approx \frac{3\rho(E_f)}{2} g(E_f, J_i), \notag
\end{align}
which is a good approximation except for the case of $J_i=0$ (and $J_i=\onehalf$), where the selection rules allow only transitions to $J_f=1$ ($J_i=\{\nicefrac{1}{2}, \nicefrac{3}{2}\} $) states.

Next, we write $n(i, f)\mathrm{d}E_f$ more explicitly as $I(E_i, J^\pi_i, E_\gamma)$ and exploit probability conservation,
\begin{align}
    \label{eq:probability_normalisation 1J}
    &P_{J^\pi_i}(E_i, E_\gamma) = \frac{I(E_i, J^\pi_i, E_\gamma)}{\sum_{E_\gamma} I(E_i, J^\pi_i, E_\gamma)} \\
    &= C_{E_i, J^\pi_i} E_\gamma^3 f_1(E_\gamma) \rho(E_i-E_\gamma) g(E_i-E_\gamma, J_i)
    \nonumber
\end{align}
where $P_{J_i, \pi_i}(E_i, E_\gamma)$ is the probability to decay from an initial excitation energy bin $E_i$ with a $\gamma$ ray of energy $E_\gamma$, the subscripts $J_i$ and $\pi_i$ limit the initial levels to of a given spin and parity, respectively, and $C_{E_i, J^\pi_i}$ is a normalization constant. Note that the normalization constant $C'_{\alpha, i}$  cancels out (which includes compound nucleus cross-section $\sigma_\mathrm{CN}$, the total radiative width $\Gamma_\gamma$, and the density of intrinsic levels $\rho(E_i, J^\pi_i)$).

The final step is to generalize this equation for the case where levels of different spins and parities are populated. Naively, we may just sum  over the decays $P_{J^\pi_i}$ from all initial levels $J^\pi_i$
\begin{align}
    \label{eq:oslo eq naive}
    \sum_{J^\pi_i} P_{J^\pi_i} & \approx C_{E_i} E_\gamma^3 f_1(E_\gamma) \rho(E_i-E_\gamma) \sum_{J^\pi_i} g(E_i-E_\gamma, J_i) \notag \\
    & = C_{E_i} E_\gamma^3 f_1(E_\gamma) \rho(E_i-E_\gamma),
\end{align}
from which we would already recover the standard Oslo method equation for a suitable normalization constant $C_{E_i}$. However, we have to note that the normalization constants $C_{E_i, J^\pi_i}$ in Eq.\ \eqref{eq:probability_normalisation 1J} depend on the spin and parity, and cannot be factored out. As we will see, this can be solved under the assumption that the cross-section $\sigma_\mathrm{CN}$ to create the compound nucleus at different spin-parities $J^\pi_i$ is proportional to number of levels in the nucleus (i.e. it is not spin-selective, but proportional to intrinsic spin distribution),
\begin{align}
    \sigma_\mathrm{CN}(\alpha \to E_i, J^\pi_i) & \approx \sigma_\mathrm{CN}(E_i) \rho(E_i, J_i, \pi_i) \notag \\
    & = \sigma_\mathrm{CN}(E_i) \rho(E_i) g(E_i, J_i, \pi_i).
\end{align}
Using Eq.\ \eqref{eq:spectrum beta simple} to \eqref{eq:sum final states}, the (cross-section weighted) sum over the decay spectra from all populated levels is
\begin{align}
    \label{eq:spectrum include Jpi}
    I(E_x, E_\gamma) & = \sum_{J^\pi_i} I(E_x, E_\gamma, J^\pi_i) \notag \\
    &= \sum_{J^\pi_i} \frac{\sigma_\mathrm{CN}(E_i) \rho(E_i, J_i, \pi_i)}{\Gamma_\gamma\rho(E_i, J_i, \pi_i)}
       f_1(E_\gamma) E_\gamma^3 \rho_\mathrm{avail}(E_f) \notag \\
    &\approx \frac{\sigma_\mathrm{CN}(E_i)}{\Gamma_\gamma} f_1(E_\gamma) E_\gamma^3 \frac{3\rho(E_f)}{2} \sum_{J^\pi_i} g(E_x, J_i) \notag \\
    &= 3\frac{\sigma_\mathrm{CN}(E_i)}{2 \Gamma_\gamma} f_1(E_\gamma) E_\gamma^3 \rho(E_x-E_\gamma).
\end{align}
In principle, we also have to average over the excitation energy bin $E_x$. However, as the level density $\rho$ and the total average radiative width $\Gamma_\gamma$ are assumed to vary only slowly with energy, this will not lead to any changes in the equation above. The normalized spectrum is given by
\begin{align}
  \label{eq:oslo eq final}
  P(E_x, E_\gamma) &= \frac{I(E_x, E_\gamma)}{\sum_{E_\gamma} I(E_x, E_\gamma)} \notag \\
  & = C_{E_x} f_1(E_\gamma) E_\gamma^3 \rho(E_f=E_x-E_\gamma),
\end{align}
for a normalization constant $C_{E_x}$ that depends only on the excitation energy.

In the case of a more spin-selective population of the compound nucleus, like in $\beta$-decay one receives a weighted sum of the level densities. If we denote the normalized population of the levels per for each excitation energy by $g_\mathrm{pop}(E_x, J, \pi)=\sigma_\mathrm{CN}(E_x, J, \pi)/\sum_{J^\pi} \sigma_\mathrm{CN}(E_x, J, \pi)$, Eq.\ \eqref{eq:spectrum include Jpi} can be generalized as
\begin{align}
  \label{eq:oslo eq weighted}
  &P(E_x, E_\gamma) \\
  &\quad = C_{E_x} f_1(E_\gamma) E_\gamma^3 \rho(E_f=E_x-E_\gamma) \sum_{J^\pi} \frac{g_\mathrm{pop}(E_x, J, \pi)}{g(E_x, J)}. \notag
\end{align}

In summary, we have shown that the Oslo method equation arises naturally from the Bohr's independence hypothesis for the compound nucleus under the assumption of i) the dominance of dipole radiation, ii) parity equilibration of the level density and iii) a compound nucleus cross-section $\sigma_\mathrm{CN}$ that is proportional to the spin distribution of intrinsic levels.

\section{Calculation of $\langle \Gamma_\gamma \rangle$}
\label{app:GammaGamma}
In this appendix we derive Eq.\ \eqref{eq:GammaGamma} used to calculate the average total radiative width $\langle \Gamma_\gamma \rangle$ from the level density $\rho$ and $\gamma$-ray strength function $f$. The derivation is based on Eqs.\ (7.19-7.23) in Ref.\ \cite{Blatt1952} and p.\,106 in Ref.\ \cite{Axel1962}. It bases on the same arguments as \ref{app:oslomethodeq}, but for ease of readability, we reiterate the main points.

The radiative width $\Gamma_\gamma$ denotes the probability for a state to decay by $\gamma$-ray emission. In the multipolar expansion it is written as
\begin{align}
    \label{eq:multipole_expansion}
    \Gamma_\gamma = \sum_L (\Gamma_{EL} + \Gamma_{ML}),
\end{align}
where $EL$ and $ML$ denote the electric and magnetic components of the radiation of order $L$, respectively. To simplify the derivation, we will now assume the dominance of dipole radiation, such that
\begin{align}
\label{eq:multipole_expansion_dipole}
\Gamma_\gamma = \Gamma_{E1} + \Gamma_{M1}.
\end{align}
We will continue with the derivations for the $E1$ radiation, but similar equations hold for $M1$. Analogously to Eq.\ \eqref{eq:spectrum beta}, we can express $\Gamma_{E1}$ as the sum of the partial decay widths $\Gamma_{E1, i \rightarrow f}$ from the initial level $i$ to allowed final levels $f$,
\begin{align}
    \label{eq:GgE1_from_partial}
    \Gamma_{E1} = \Gamma_{\gamma,E1} (E_x, J_i, \pi_i) =\sum_f \Gamma_{E1,i \to f},
\end{align}
where the initial level is at the excitation energy $E_x$ and has the spin $J_i$ and parity $\pi_i$. The angular momentum and parity of the final states are given by the usual selection rules.

We now rewrite the partial widths $\Gamma_{E1,i \to f}$ in terms of the strength function $f$, using the definition of $f$, Eq.\ \eqref{eq:gsf_definition}, and the generalized Brink-Axel hypothesis, Eq.\ \eqref{eq:gsf_Brink_Axel},
\begin{align}
    \label{eq:gsf-definition}
    \Gamma_{E1,i \to f} \to \langle \Gamma_{E1}(E_\gamma, E_x) \rangle = \frac{f_{E1}(E_\gamma) E_\gamma^3}{\rho(E_x, J_i, \pi_i)}
\end{align}
where $\rho(E_x, J_i, \pi_i)$ is the density of the initial levels, and the energy difference to the final state(s) is given by $E_\gamma=E_x-E_f$.

Next, we replace the the sum in Eq.\ \eqref{eq:GgE1_from_partial} by an integral, where we note that the number of partial widths $\langle \Gamma_{E1}(E_\gamma, E_x) \rangle$ is proportional to the level density at the final states
\begin{align}
\sum_{J_f}\sum_{\pi_f}\rho(E_x - E_\gamma, J_f, \pi_f),
\end{align}
such that the average total radiative width $\langle \Gamma_{\gamma,E1} (E_x, J_i, \pi_i) \rangle$ is given by
\begin{align}
  \label{eq:appGammaGamma_1}
   & \langle \Gamma_{\gamma,E1} (E_x, J_i, \pi_i)\rangle \\
   &  = \int_0^{E_x} dE_\gamma\, \langle \Gamma_{E1}(E_\gamma, E_x) \rangle \sum_{J_f} \sum_{\pi_f} \rho(E_x - E_\gamma, J_f, \pi_f) \notag \\
   &  = \int_0^{E_x} dE_\gamma\, \frac{f_{E1}(E_\gamma) E_\gamma^3}{\rho(E_x, J_i, \pi_i)} \sum_{J_f=|J_i-1|}^{J_i+1} \rho(E_x-E_\gamma, J_f, \bar\pi_i)  \notag
\end{align}
where the $E1$ operator requires a change of parity for the final state, here denoted as $\bar\pi_i$. The same selection rules as given in footnote \ref{fn:selection rules} have been applied, but we do not use the approximation of Eq.\ \eqref{eq:sum final states}, as many nuclei have a initial spin $J_i$ of 0 or $\onehalf$. Further, we assume parity equilibration of the level density, i.e.\ ~$\rho(E_x, J, +) \approx \rho(E_x, J, -)$, and express the level density $\rho(E_x, J, \pi)$ through the spin-distribution $g(E_x, J)$, Eq.\ \eqref{eq: spin dist eq},
\begin{align}
  \label{eq:appGammaGamma_2}
   & \langle \Gamma_{\gamma,E1} (E_x, J_i, \pi_i)\rangle = \frac{1}{\rho(E_x, J_i, \pi_i)} \\
   & \times \int_0^{E_x} dE_\gamma\, \Biggl [f_{E1}(E_\gamma) E_\gamma^3  \rho(E_x - E_\gamma) \sum_{J_f=|J_i-1|}^{J_i+1} \frac{g(E_x-E_\gamma, J_f)}{2} \Biggr ].  \notag
\end{align}
At this point, to obtain the expression for $M1$ radiation, one only needs to exchange the $E1$ strength function $f_{E1}$ by the $M1$ strength function $f_{M1}$.  Using Eq.\ \eqref{eq:multipole_expansion_dipole}, we find
\begin{align}
  \label{eq:appGammaGamma_3}
   & \langle \Gamma_{\gamma} (E_x, J_i, \pi_i)\rangle = \frac{1}{2\rho(E_x, J_i, \pi_i)} \\
   & \times \int_0^{E_x} dE_\gamma\, \Biggl [\bigl[f_{E1}(E_\gamma)+f_{M1}(E_\gamma)\bigr] E_\gamma^3  \rho(E_x - E_\gamma) \notag\\
   & \times  \sum_{J_f=|J_i-1|}^{J_i+1} g(E_x-E_\gamma, J_f) \Biggr ].  \notag
\end{align}
This is equivalent to Eq.\ (2.11) in Ref.\ \cite{Bartholomew1973} under the assumptions listed above. One can determine $\langle \Gamma_{\gamma} (E_x, J_i, \pi_i)\rangle$ from neutron capture experiments, where usually slow neutrons are used, thus $E_x \approx S_n$. This is shown e.g.\ in Ref.\ \cite{Voinov2001}, but we will repeat the derivation to get a comprehensive picture. The intrinsic spin of the neutron is $\onehalf$,
so with capture of order $\ell$ on a target (denoted with the subscript $t$) the entry states in the residual nucleus have the possible spins $J_i = J_t \pm \onehalf \pm \ell$ and parity $\pi_i = \pi_t (-1)^\ell$. For s-wave capture on a target nucleus with $J_t = 0$, there is only one possible $J_i$, and we can directly compare the experimental measurements to the calculations using \eqref{eq:appGammaGamma_3}. For $J_t > 0$, often only the average over all resonance of different $J_i$ is reported. Using the level density $\rho(S_n, J_i)$ of the accessible $J_i$'s, we find
\begin{align}
  \label{eq:appGammaGamma_4}
   & \langle \Gamma_{\gamma \ell} (S_n) \rangle = \frac{\sum_{J_i} \rho(S_n, J_i) \langle \Gamma_{\gamma}(S_n, J_i, \pi_i)}{\sum_{J_i} \rho(S_n, J_i)} \\
   & = \frac{D_l}{2} \int_0^{S_n} dE_\gamma\, \Biggl [f(E_\gamma) E_\gamma^3  \rho(S_n - E_\gamma) \notag\\
   & \quad \times  \sum_{J_i} \sum_{J_f=|J_i-1|}^{J_i+1} g(S_n-E_\gamma, J_f) \Biggr ],  \notag
\end{align}
where we substituted $\sum_{J_i} \rho(S_n, J_i)$ by the average neutron resonance spacing $D_l$ and used the assumption of the dominance of the dipole decay to rewrite the sum of the average $M1$ and $E1$ strengths as $f(E_\gamma$). Note that transitions for the highest $\gamma$-ray energies go to discrete states, not a quasi(-continuum). Thus, this integral is an approximation, and more precise calculations could distinguish between a sum over transitions to discrete states and the integral for the (quasi-)continuum region.

\section*{References}
\bibliography{bib}

\end{document}